\documentclass[12pt,letterpaper]{JHEP3}

\usepackage{amscd,amsmath,amssymb,amsfonts,xspace,mathrsfs}
\usepackage{epsfig}

\def\lfig#1#2#3#4#5{
 \begin{figure}
 \refstepcounter{figure}
 \label{#4}
 \addtocounter{figure}{-1}
 \epsfxsize=#3
 \centerline{\epsfbox{#2}}
 \vspace{#5}
 {\bf \caption{{\rm #1}}}
 \end{figure}
}

\numberwithin{equation}{section}

\def\varpi{t}

\def\sign{{\rm sign}}

\def\Re{\,{\rm Re}\,}

\def\({\left(}
\def\){\right)}
\def\[{\left[}
\def\]{\right]}
\def\hf{{1\over 2}}

\newcommand{\de}{\mathrm{d}}

\newcommand{\I}{\mathrm{i}}

\newcommand{\cL}{\mathcal{L}}
\newcommand{\cD}{\mathcal{D}}

\def\vrh{\varrho}

\newcommand{\p}{\partial}

\newcommand{\cV}{\mathcal{V}}
\newcommand{\cC}{\mathcal{C}}

\newcommand{\cM}{\mathcal{M}}

\newcommand{\cE}{\mathcal{E}}
\newcommand{\cX}{\mathcal{X}}
\newcommand{\CX}{\mathcal{X}}

\newcommand{\unit}{{\mathbf{1}}}

\DeclareSymbolFont{AMSa}{U}{msa}{m}{n}
\DeclareSymbolFont{AMSb}{U}{msb}{m}{n}
\DeclareMathSymbol{\fieldR}{\mathalpha}{AMSb}{"52}

\newcommand{\kahler}{{K\"ahler}\xspace}

\newcommand{\qk}{{quaternion-K\"ahler}\xspace}

\newcommand{\cZ}{\mathcal{Z}}

\newcommand{\cO}{\mathcal{O}}

\newcommand{\cU}{\mathcal{U}}
\newcommand{\cA}{\mathcal{A}}

\newcommand{\pa}{\partial}

\newcommand{\eps}{\epsilon}
\newcommand{\IR}{\mathbb{R}}
\newcommand{\IC}{\mathbb{C}}
\newcommand{\IZ}{\mathbb{Z}}

\newcommand{\IP}{\mathbb{P}}

\newcommand{\tzeta}{\tilde\zeta}

\newcommand{\txi}{\tilde\xi}

\newcommand{\CP}{\IC P^1}

\def\bea{\begin{eqnarray}}
\def\eea{\end{eqnarray}}
\def\be{\begin{equation}}
\def\ee{\end{equation}}
\def\ba{\begin{align}}
\def\ea{\end{align}}
\def\bse{\begin{subequations}}
\def\ese{\end{subequations}}

\def\ba{\bar a}

\def\bF{\bar F}

\def\htt{{\mathtt t}}
\def\htp{\htt_+}
\def\htm{\htt_-}
\def\htpm{\htt_\pm}

\def\tS{\tilde S}

\newcommand{\cB}{\mathcal{B}}
\def\hH{h}

\def\cij#1{c}
\def\ci#1{c}

\def\bn{\bar n}

\def\ui#1{^{[#1]}}

\def\txii#1{{\tilde\xi}^{[#1]}}

\def\ai#1{{\alpha}^{[#1]}}

\def\xii#1{\xi_{[#1]}}

\def\gi#1{g^{[#1]}}

\def\Hij#1{H^{[#1]}}

\newcommand{\Li}{{\rm Li}}

\def\tleta{\tilde\eta}

\def\XXint#1#2#3{{\setbox0=\hbox{$#1{#2#3}{\int}$}
\vcenter{\hbox{$#2#3$}}\kern-.5\wd0}}

\newcommand{\cwarrow}{\text{\Large$\curvearrowright$}}
\newcommand{\ccwarrow}{\text{\Large$\curvearrowleft$}}

\def\muh{\mu}

\def\hHij#1{\hH^{[#1]}}
\def\hkp{\hHij{\hgam}_{k,p}}

\def\gfinv{n^{(0)}}

\def\Fcl{F^{\rm cl}}
\def\bFcl{\bF^{\rm cl}}

\newcommand{\expe}[1]{{\bf E}\!\left( #1\right)}

\def\ws{{\rm w.s.}}

\def\gamD#1{\tilde\gamma}
\def\GamD#1{\Gamma^{(#1)}}

\def\CY{\mathfrak{Y}}
\def\CYm{\mathfrak{Y}}

\def\gl#1{{\rm g}_{#1}}

\def\Nq{N}

\DeclareMathOperator{\Td}{Td}
\DeclareMathOperator{\ch}{ch}

\def\cla{\tilde c_a}
\def\cl0{\tilde c_0}

\def\thetaD{\theta_{\rm D}}

\def\hgam{\hat \gamma}

\def\SK{\cM_{\rm ks}}

\def\qfD#1{\sigma_D(#1)}

\def\Xf#1{X_{#1}}

\def\ii{i}

\def\indlp{\text{in}}
\def\indrp{\text{out}}

\def\Tkp#1{T^{[\hgam]\,#1}_{k,p}}
\def\tTkp#1{\tilde T^{[\hgam]}_{k,p;\,#1}}
\def\tTkpa{\tilde T^{[\hgam]}_{k,p;\,\alpha}}
\def\Tmn#1{T^{#1}}

\def\opT{\mathbf{T}}
\def\opTH{T}

\title{Dualities and fivebrane instantons}

\author{Sergei Alexandrov and Sibasish Banerjee
\\
{\it Universit\'e Montpellier 2, Laboratoire Charles Coulomb UMR 5221, F-34095,
Montpellier, France}\\

\vspace*{2mm} {\tt e-mail:
\email{salexand@univ-montp2.fr},
\email{sibasishbanerjee@live.in}
}

\vspace*{-3mm}

}

\abstract{We derive the fivebrane instanton corrections to the hypermultiplet moduli space $\cM_H$
of Calabi-Yau string vacua using S-duality symmetry of the type IIB formulation.
The result is given in terms of a set of holomorphic functions on the twistor space of $\cM_H$.
It contains not only all orders of the instanton expansion, but also takes into account the presence of D1-D(-1)-brane instantons.
Furthermore, we provide a thorough study of the group of discrete isometries of $\cM_H$ and show
that its closure requires a modification of certain symmetry transformations.
After this modification, the fivebrane instantons are proven to be consistent with the full duality group.
}

\begin{document}

\section{Introduction}

Compactifications of type II string theory on Calabi-Yau (CY) threefolds represent a fruitful laboratory to
generate, test, and exemplify various ideas on string dynamics, dualities and non-perturbative physics.
They are very rich from both, physical and mathematical, points of view and have numerous relations with other
subjects such as BPS black holes, supersymmetric gauge theories, integrable systems, etc.
Moreover, in contrast to compactifications with fewer preserved supersymmetries,
CY vacua seem to be amenable for an exact description. Although such a description, which is supposed to provide
the complete {\it non-perturbative} low energy effective action for the compactification on arbitrary CY $\CY$,
has not been achieved yet, this goal appears now within our reach.

Let us summarise what is known about this problem up to now (see \cite{Alexandrov:2011va,Alexandrov:2013yva} for reviews).
At two derivative level, the low energy action is completely determined by the metrics on the moduli spaces of vector multiplets (VM)
and hypermultiplets (HM), $\cM_V$ and $\cM_H$ \cite{deWit:1984px}. The former is a special \kahler manifold whose geometry is determined
by a holomorphic prepotential $F(X)$, a homogeneous function of degree 2, which is in principle known for arbitrary CY
in terms of its topological data \cite{Candelas:1990rm,Hosono:1993qy}:
triple intersection numbers $\kappa_{abc}$, Euler characteristic $\chi_\CY$,
and genus zero Gopakumar--Vafa invariants $\gfinv_{q_a}$.
On the other hand, the latter is a \qk (QK) manifold \cite{Bagger:1983tt},
receiving stringy quantum corrections, whose exact geometry is not known yet and represents the main challenge.

The quantum $g_s$-corrections to the classical metric on $\cM_H$ can be split into perturbative and non-perturbative ones,
and the latter come either from (Euclidean) D-branes wrapping non-trivial cycles of the CY,
or from NS5-branes wrapped on the whole compactification manifold \cite{Becker:1995kb}.
Remarkably, only the very last set of corrections, namely those given by NS5-brane instantons, remain unknown so far.
More precisely, the perturbative corrections are restricted to one-loop and have been incorporated in
\cite{Antoniadis:1997eg,Gunther:1998sc,Antoniadis:2003sw,Robles-Llana:2006ez,Alexandrov:2007ec}.
All D-instantons have been described in \cite{Alexandrov:2008gh,Alexandrov:2009zh} within the type IIA formulation.
Finally, in \cite{Alexandrov:2010ca} an attempt to include NS5-instantons in the one-instanton approximation has been made
using the mirror type IIB framework.

As a result, what remains is to find NS5-brane corrections beyond the one-instanton approximation.
This is precisely the goal of the present paper. In fact, we have already announced our main results in a short note \cite{Alexandrov:2014mfa}.
Here we provide their detailed derivation and extend them by including the effects of D1-D(-1)-instantons.

More precisely, we concentrate on the type IIB formulation where all quantum corrections to the metric on $\cM_H$
can be arranged into sectors invariant under the action of the S-duality group $SL(2,\IZ)$.
This can be represented by the following table:
\be
\mbox{
\begin{tabular}{l|c|c|c|c|c|c|cc|}
\cline{2-2} \cline{4-4}
$\alpha'$-corrections: \hspace{0.1cm} & perturbative & \hspace{0.1cm} & w.s. instantons  & \multicolumn{4}{c}{} \rule{0pt}{12pt}
\\
\cline{6-6} \cline{8-9}
$g_s$-corrections: & 1-loop \ \ D(-1) & & D1 & \hspace{0.1cm} & \,D3\, & \hspace{0.1cm} & \,D5 &  NS5 \rule{0pt}{13pt}
\\
\cline{2-2} \cline{4-4} \cline{6-6} \cline{8-9}
\end{tabular}}
\label{quantcor}
\ee
and makes possible studying each sector independently of the others. Moreover, one can use S-duality
to find all quantum corrections inside some sector if one knows already at least a part of them.
It is sufficient just to apply the method of images. For instance,
this was precisely the idea used in \cite{RoblesLlana:2006is}
to find D1 and D(-1)-instantons from the knowledge of $\alpha'$-corrections
encoded in the holomorphic prepotential $F(X)$.
Looking at the pattern \eqref{quantcor}, it is tempting to apply the same idea to the last sector to obtain NS5-instanton corrections
from D5-instantons, which follow from the results of \cite{Alexandrov:2008gh,Alexandrov:2009zh} and mirror symmetry.
This was realized in \cite{Alexandrov:2010ca}, but only in the one-instanton approximation due to
several complications arising on the way.

The first difficulty is related to the action of S-duality.
As we will review below, instanton corrections to the HM moduli space have the simplest incarnation
in the twistor space $\cZ$ of $\cM_H$, and are encoded in a set of holomorphic functions, known as {\it transition functions}.
Therefore, to derive NS5-instantons from D5 ones, it is important to know how S-duality acts on the transition functions.
This was understood only recently in \cite{Alexandrov:2013mha} and, unfortunately, the resulting action
turned out to be highly non-linear which makes its application very non-trivial.

The second complication is that the sectors in \eqref{quantcor} are not actually completely independent. As we will see,
when translating the results on D-instanton corrections from type IIA to the manifestly S-duality invariant framework,
adapted to the symmetries of the type IIB formulation, the first three sectors affect the last one.
Thus, this effect should be taken into account in the complete picture including all quantum corrections.

In this paper we show how both these difficulties can be overcome.
A way to avoid the first one was in fact already proposed in \cite{Alexandrov:2014mfa}, and is based on an alternative
parametrization of the twistor space which uses, instead of the usual transition functions, certain {\it contact Hamiltonians}.
This allows to linearize the action of S-duality so that the derivation of fivebrane instantons becomes straightforward.
Here we also include into this description the effects of D1-D(-1)-instantons coming from the first two sectors in \eqref{quantcor}.

Thus, we provide the twistorial formulation of the non-perturbative geometry of $\cM_H$ where only D3-instantons
are missing. Although they are known on the mirror type IIA side, where they appear as a subset of D2-brane instantons,
their manifestly S-duality invariant formulation, which is what we really need here, has not been
found yet.\footnote{The work in this direction was initiated in \cite{Alexandrov:2012au} where it was shown that
the type IIA construction of these instanton corrections is consistent with S-duality at least in the one-instanton approximation.
However, the corresponding twistorial formulation adapted to this symmetry is still lacking.}
This is related to the fact, distinguishing them from other instanton corrections and clearly seen from \eqref{quantcor},
that they are {\it selfdual} under $SL(2,\IZ)$. Thus, a better understanding of these instanton corrections is required before
including them into our picture.

Another important result, which we present here, is an improved understanding of the discrete isometry group of $\cM_H$.
Already in \cite{Alexandrov:2010ca} it was observed that the fivebrane corrections obtained by applying S-duality as described above
appear to be incompatible with other discrete symmetries such as large gauge transformations of the RR-fields and
monodromy transformations of the complexified \kahler moduli. We trace this incompatibility back to
the failure of the generators of these discrete isometries to form a group representation.
At the same time, we show how this situation can be cured by adjusting the action of monodromies on the RR-scalars
and demonstrate that our results on fivebrane instantons are consistent with the resulting duality group.

The organisation of the paper is as follows. In the next section we present the basic information about
the HM moduli space concentrating on the type IIB formulation. Here we also discuss the isometries of $\cM_H$, the subtleties
related to their action at quantum level, and provide the corrected form of the discrete symmetry transformations.
In section \ref{sec-twistor} we review the twistorial construction of QK manifolds, improved parametrization introduced
in \cite{Alexandrov:2014mfa}, and constraints imposed by the presence of the $SL(2,\IZ)$ isometry group.
In section \ref{sec-Dinst} this twistor framework is used to describe D-instanton corrections, after which it is shown
how D1-D(-1)-instantons can be reformulated in a manifestly S-duality invariant way and how this reformulation affects
other D-instanton contributions. Then in section \ref{sec-fivebrane} we derive the fivebrane instantons at all orders
in the instanton expansion. Section \ref{sec-concl} present our conclusions.
In addition, in appendix \ref{ap-Udual} we provide details on the isometry group of $\cM_H$.
In appendix \ref{ap-contactbr} we give a proof of a crucial transformation property of our twistorial construction.
Appendix \ref{ap_Sdualconstrap} verifies that the non-linear S-duality constraint of \cite{Alexandrov:2013mha}
is indeed satisfied by the transition functions of fivebrane instantons which we compute in this paper.
In appendix \ref{ap-MHinv} we check that the twistorial construction of fivebrane instantons is compatible
with all isometries expected to survive quantum corrections.
And finally, in the last appendix we provide explicit expressions for derivatives of fivebrane transition functions.
They are to be used in the integral equations determining the metric on $\cM_H$ which includes all quantum corrections except D3-instantons.

\section{Hypermultiplet moduli space in CY compactifications}
\label{sec-HM}

\subsection{Classical moduli space}

In this section we review the main facts about the hypermultiplet moduli space $\cM_H$ of CY string vacua,
with emphasis on its symmetries at classical and quantum level.
This moduli space appears in the two versions corresponding to type IIA and type IIB formulations of string theory,
but mirror symmetry, or more precisely its non-perturbative extension \cite{Ferrara:1995yx},
requires them to coincide if the compactification manifolds in the two formulations are chosen to be mirror to each other.
Here we will mostly work with the type IIB version since it is better suited to the application of S-duality.

In type IIB string theory compactified on a CY threefold $\CYm$, $\cM_H$ is a QK manifold of real
dimension $4 (h_{1,1} (\CYm) + 1)$. It comes with a set of natural coordinates which correspond
to scalar fields in four dimensions and comprise
\begin{itemize}
\item the ten-dimensional dilaton equal to the inverse string coupling $\tau_2=1/g_s$;
\item the K\"ahler moduli $b^a + \I t^a\equiv \int_{\gamma^a} \mathcal{J}$
($a=1,\dots, h_{1,1}$)
where $\mathcal{J} \equiv B+\I\, J$ is the complexified \kahler form on  $\CYm$
and $\gamma^a$ is a basis of  $H_{2}(\CYm,\IZ)$;
\item the  Ramond-Ramond (RR) scalars $c^0,c^a,\cla,\cl0$, corresponding to (suitable combinations of) periods of the RR 0-form,
2-form, 4-form and 6-form potentials;
\item the NS axion $\psi$, dual to the 2-form $B$ in four dimensions.
\end{itemize}
It is useful also to combine the string coupling and the RR scalar $\tau_1 = c^0$ into an
axio-dilaton field $\tau = \tau_1 + \I \tau_2$.

At tree level the metric on $\cM_H$ is given by the so-called {\it local c-map} \cite{Ferrara:1989ik}.
We do not need its explicit expression in this paper. What is important for us is that it is completely determined
by the holomorphic prepotential on the \kahler structure moduli space $\SK$ of $\CYm$.
The prepotential is known to have the following form \cite{Candelas:1990rm,Hosono:1993qy}
\be
\label{lve}
F(X)=-\kappa_{abc}\frac{X^a X^b X^c}{6 X^0}
+ \chi_{\CYm}\frac{\zeta(3)(X^0)^2}{2(2\pi\I)^3}
-\frac{(X^0)^2}{(2\pi\I)^3}{\sum_{q_a\gamma^a\in H_2^+(\CYm)}} \gfinv_{q_a}\,
\Li_3\[ \expe{q_a\, \frac{X^a}{X^0}}\],
\ee
where $X^\Lambda$ ($\Lambda=0,\dots,h_{1,1}$) are homogeneous coordinates related to the \kahler moduli by
$X^a/X^0 = b^a + \I t^a$ and we introduced the convenient notation $\expe{x}=e^{2\pi\I x}$.
In \eqref{lve} the first term describes the classical part of the prepotential, whereas the second and third terms
correspond to a perturbative $\alpha'$-correction and contributions of worldsheet instantons, respectively.
The instantons are labeled by effective homology classes $q_a\gamma^a\in H_2^+(\CYm)$, which means that $q_a\ge 0$ for all $a$,
not all of them vanishing simultaneously, and introduced via the trilogarithm function $\Li_3(x)=\sum_{n=1}^\infty x^n/n^3$.

It is useful also to introduce another set of coordinates which appears to be more convenient in the mirror type IIA formulation.
The relation between the two coordinate sets is known as the {\it classical mirror map}
\cite{Bohm:1999uk}
\be
\label{symptobd}
\begin{split}
z^a & =b^a+\I t^a\, ,
\qquad\ \
\zeta^0=\tau_1\, ,
\qquad\
\zeta^a = - (c^a - \tau_1 b^a)\, ,
\\
\tzeta_a &=  \cla+ \frac{1}{2}\, \kappa_{abc} \,b^b (c^c - \tau_1 b^c)\, ,
\qquad
\tzeta_0 = \cl0-\frac{1}{6}\, \kappa_{abc} \,b^a b^b (c^c-\tau_1 b^c)\, ,
\\
\sigma &= -2 (\psi+\frac12  \tau_1 \cl0) + \cla (c^a - \tau_1 b^a)
-\frac{1}{6}\,\kappa_{abc} \, b^a c^b (c^c - \tau_1 b^c)\, .
\end{split}
\ee
Using the type IIA coordinates, we can easily write down the continuous transformations leaving the tree level metric on $\cM_H$
invariant. These are the so-called Peccei-Quinn symmetries arising due to the fact that the RR-scalars and the NS-axion
originate from gauge fields.
They act by shifting the corresponding scalars and form the Heisenberg group
\be
\label{heis0}
\opTH_{\eta^\Lambda,\tleta_\Lambda,\kappa}\ :\quad
\bigl(\zeta^\Lambda,\tzeta_\Lambda,\sigma\bigr)\ \mapsto\
\bigl(\zeta^\Lambda + \eta^\Lambda ,\
\tzeta_\Lambda+ \tleta_\Lambda,\
\sigma + 2 \kappa- \tleta_\Lambda \zeta^\Lambda
+ \eta^\Lambda \tzeta_\Lambda  \bigr).
\ee
Furthermore, in the large volume limit, where one can drop the last two terms in the prepotential \eqref{lve},
there are additional symmetries. One of them is another Peccei-Quinn symmetry shifting the scalars $b^a$
coming from the 2-form gauge field $B$. This shift however should be accompanied by certain transformations
of the RR-scalars so that the full transformation is given by
\be
\label{bjacr}
M_{\epsilon^a}\ :\quad  \begin{array}{c}
\displaystyle{b^a\mapsto b^a+\epsilon^a\, ,
\qquad
\zeta^a\mapsto \zeta^a + \epsilon^a \zeta^0\, ,
\qquad
\tzeta_a\mapsto \tzeta_a -\kappa_{abc}\zeta^b \epsilon^c
-\frac12\,\kappa_{abc} \epsilon^b \epsilon^c \zeta^0\, ,}
\\
\displaystyle{\tzeta_0\mapsto \tzeta_0 -\tzeta_a \epsilon^a+\frac12\, \kappa_{abc}\zeta^a \epsilon^b \epsilon^c
+\frac16\,\kappa_{abc} \epsilon^a \epsilon^b \epsilon^c \zeta^0\, .}
\end{array}
\ee
And finally the classical metric in the large volume limit is invariant under transformations which
form the $SL(2,\IR)$ group and, in contrast to the previous ones, are most easily written in the type IIB field basis
\be\label{SL2Z}
SL(2,\IR)\ni\gl{}\ :\quad
\begin{array}{c}
\displaystyle{
\tau \mapsto \frac{a \tau +b}{c \tau + d} \, ,
\qquad
t^a \mapsto t^a |c\tau+d| \, ,
\qquad
\cla\mapsto \cla \,  ,}
\\
\displaystyle{
\begin{pmatrix} c^a \\ b^a \end{pmatrix} \mapsto
\begin{pmatrix} a & b \\ c & d  \end{pmatrix}
\begin{pmatrix} c^a \\ b^a \end{pmatrix}\, ,
\qquad
\begin{pmatrix} \cl0 \\ \psi \end{pmatrix} \mapsto
\begin{pmatrix} d & -c \\ -b & a  \end{pmatrix}
\begin{pmatrix} \cl0 \\ \psi \end{pmatrix},}
\end{array}
\ee
with $ad-bc=1$. As we review below, all these continuous isometries are lifted by quantum corrections, but
at the same time each of them leaves an unbroken discrete subgroup.

\subsection{Quantum corrections}

Besides the $\alpha'$-corrections completely captured by the prepotential \eqref{lve}, the HM moduli space receives
$g_s$-corrections. At perturbative level, there is only a one-loop correction controlled by the Euler characteristic $\chi_{\CYm}$.
The resulting metric is a one-parameter deformation of the c-map metric whose explicit form can be found in \cite{Alexandrov:2007ec}.

The situation is more interesting at the non-perturbative level where one finds two types of instanton contributions.
The first type comes from D-branes wrapping non-trivial cycles of the CY compactification manifold and has
the following generic form
\be
\label{d2quali}
\delta \de s^2\vert_{\text{D-inst}} \sim \qfD{\gamma}\,\Omega(\gamma;z)\,
 e^{ -2\pi|Z_\gamma|/g_s
- 2\pi\I (q_\Lambda \zeta^\Lambda-p^\Lambda\tzeta_\Lambda)} .
\ee
Here $\gamma=(p^\Lambda,q_\Lambda)$ is the D-brane charge, the function $Z_\gamma(z)$ is the the central charge of the supersymmetry
subalgebra preserved by the instanton, which is given by ($z^0\equiv 1$)
\be
\label{defZ}
Z_\gamma(z) = q_\Lambda z^\Lambda- p^\Lambda F_\Lambda(z),
\ee
$\Omega(\gamma;z)$ are generalized Donaldson-Thomas invariants (BPS indices)
dependent of the moduli $z^a$ in a piecewise constant way,
and finally $\qfD{\gamma}$ is the so-called quadratic refinement factor whose defining property is
\be
\qfD{\gamma}\qfD{\gamma'}=(-1)^{\langle\gamma,\gamma'\rangle}\qfD{\gamma+\gamma'},
\label{defqr}
\ee
where $\langle\gamma,\gamma'\rangle=q_\Lambda p'^\Lambda-q'_\Lambda p^\Lambda$ is the Dirac-Schwinger product.

On the type IIB side, a mathematically rigorous way to think about D-instantons
is as objects in the derived category of coherent sheaves ${\rm D^{b}Coh}(\CYm)$ \cite{Sharpe:1999qz,Douglas:2000ah}.
Then the charge is given by the generalized Mukai vector
\be
\gamma= \ch (\mathscr{E}) \, \sqrt{\Td \CYm}
= p^0 + p^a \omega_a - q_a \omega^a + q_0\, \omega_{\CYm}\, ,
\label{chMu}
\ee
where $\mathscr{E}$ is a coherent sheaf, and $\{\omega_a\}$, $\{\omega^a\}$, $\omega_{\CYm}$ are
respectively a basis of 2-forms, 4-forms and the volume form of $\CYm$.
For non-vanishing $p^0$ the sheaf describes a bound state of D5, D3, D1 and D(-1)-branes
with charges given by the components of $\gamma=(p^0,p^a,q_a,q_0)$.
If $p^0=0$ but $p^a$ is non-vanishing, the coherent sheaf is supported on a divisor and describes a D3-instanton, etc.
What is important is that the expression \eqref{chMu} leads to {\it non-integer} D1-D(-1)-charges $q_\Lambda$ which
satisfy the following quantization conditions
\be
\label{fractionalshiftsD5}
q_a \in \IZ - \frac{p^0}{24}\, c_{2,a} - \frac12\, \kappa_{abc} p^b p^c ,
\qquad
q_0\in \IZ-\frac{1}{24}\, p^a c_{2,a} ,
\ee
where $c_{2,a}$ are the components of the second Chern class of $\CYm$ in the basis $\omega^a$.
In other words, the charge vector is an element of $H^{\text{even}}(\CYm,\mathbb{Q})$.
On the other hand, on the type IIA side all D-brane charges are integer.
To reconcile these two facts with mirror symmetry, one should note that the holomorphic prepotential, which one obtains by applying
this symmetry, is not exactly the same as in \eqref{lve}, but differs from it by a quadratic contribution
\cite{Candelas:1990rm,Hosono:1993qy}
\be
F_{\rm m.s.}(X)=F(X)+ \frac12 \,A_{\Lambda\Sigma} X^\Lambda X^\Sigma.
\label{fullprep}
\ee
The additional term is characterized by a real symmetric matrix $A_{\Lambda\Sigma}$.
Although, as can be easily checked, it does not affect the \kahler potential of the special \kahler manifold $\SK$,
it is this term that ensures the consistency of charge quantization with mirror symmetry
and, as will be shown below, plays an important role in the correct implementation of discrete symmetries of $\cM_H$ at full quantum level.
The idea is that the type IIA and type IIB charge vectors are related by a symplectic transformation generated by $A_{\Lambda\Sigma}$.
It affects both, charges and fields,\footnote{In \cite{Alexandrov:2010ca,Alexandrov:2011va} the charges $q_\Lambda$
and the RR-fields $\tzeta_\Lambda$ were denoted by $q'_\Lambda$ and $\tzeta'_\Lambda$, respectively, whereas the unprimed notations
were reserved for the charges and fields in the type IIA frame.
However, since in this paper we work mostly in the type IIB basis, we omit the prime.
\label{foot-prime}}
\be
\tzeta_\Lambda\ \mapsto\ \tzeta_\Lambda+A_{\Lambda\Sigma}\zeta^\Lambda,
\qquad
q_\Lambda\ \mapsto\ q_\Lambda+A_{\Lambda\Sigma} p^\Sigma,
\label{symA}
\ee
and also restores the quadratic term in the prepotential \eqref{fullprep}.
It turns out that the properties satisfied by this matrix (see \eqref{propA})
are sufficient to ensure the integrality of the transformed charges \cite{Alexandrov:2010ca}.
Note that the central charge \eqref{defZ} and the whole D-instanton correction \eqref{d2quali} are symplectic invariant
and are not affected by the transformation \eqref{symA}.

The second type of non-perturbative corrections is provided by NS5-brane instantons wrapping the whole CY.
Their general form is
\be
\delta \de s^2\vert_{\text{NS5-inst}}  \sim
e^{-2 \pi |k| \cV /g_s^2+\I\pi  k \sigma},
\label{couplNS5}
\ee
where $\cV$ is the Calabi-Yau volume. In the small string coupling limit they are exponentially suppressed comparing
to the D-instantons \eqref{d2quali}. However, for finite coupling they cannot be neglected and represent an important
non-perturbative contribution.

\subsection{The duality group}
\label{subsec-Udual}

\subsubsection{Discrete isometries}
\label{subsubsec-isom}

An immediate consequence of the presence of the instanton corrections \eqref{d2quali} and \eqref{couplNS5} is that they
break the Heisenberg group of continuous transformations \eqref{heis0}. Furthermore, already the $\alpha'$-corrections
to the holomorphic prepotential break the other two continuous symmetries, \eqref{bjacr} and \eqref{SL2Z}.
Thus, the non-perturbative metric on the HM moduli space does not have {\it any} continuous isometries.

Nevertheless, each of the broken continuous groups leaves an unbroken discrete subgroup.
Before we discuss these discrete isometries, we need to provide a more detailed information on the two objects
appearing in the discussion of D-instanton corrections: the matrix $A_{\Lambda\Sigma}$ and the quadratic refinement $\qfD{\gamma}$.

The matrix $A_{\Lambda\Sigma}$ is known to satisfy the following conditions \cite{Hosono:1994av,Alexandrov:2010ca}
\be
A_{00}\in \IZ ,
\qquad
A_{0a} = \frac{c_{2,a}}{24}+ \IZ ,
\qquad
\frac12\, \kappa_{abc} \eps^b \eps^c-A_{ab}\eps^b\in \IZ \quad \text{for}\ \forall \eps^a\in\IZ.
\label{propA}
\ee
Without loss of generality, we can drop the possible integer contributions to $A_{0\Lambda}$ since they can always be removed by
an integer valued symplectic transformation. Thus, we set
\be
A_{00}=0,
\qquad
A_{0a} = \frac{c_{2,a}}{24}.
\label{valA}
\ee
An explicit expression for the components $A_{ab}$, restricted by \eqref{propA} to be half-integer,
has been found in the one modulus case in \cite{Huang:2006hq} and reads
\be
A_{11}=\hf\int_{\CYm}\iota_\star c_1(D)\wedge J,
\ee
where $D$ is the divisor dual to $J$. Although this formula begs for a generalization, it is not clear to us
how to ensure that the resulting matrix is symmetric. For most purposes, the properties listed in \eqref{propA}
turn out to be sufficient, provided they are supplemented by another property\footnote{The property \eqref{c2aprop}
follows from the fact that the expression on the l.h.s. is the holomorphic Euler characteristic of the divisor $\gamma_a$
Poincar\'e dual to  the 2-form $\omega_a$. Besides, the third condition in \eqref{propA} implies another
restriction on the intersection numbers, $\frac12\(\kappa_{aab}-\kappa_{abb}\)\in\IZ$, which in turn can be derived
from an index theorem \cite{1966InMat...1..355W}.
We thank R. Valandro for clarifying the origin of these relations.}
satisfied by the second Chern class coefficients \cite{Maldacena:1997de,Denef:2007vg}
\be
\frac16\, \kappa_{abc}\eps^a\eps^b\eps^c+\frac{1}{12}\, c_{2,a}\eps^a\in \IZ \quad \text{for}\ \forall \eps^a\in\IZ.
\label{c2aprop}
\ee

The quadratic refinement factor $\qfD{\gamma}$ typically appears in chiral boson partition functions
\cite{AlvarezGaume:1986mi,AlvarezGaume:1987vm,Witten:1996hc,Freed:2000ta}. Here it is required by consistency with the wall-crossing
to ensure the smoothness of the metric across lines of marginal stability where the BPS indices $\Omega(\gamma)$ may jump \cite{Alexandrov:2011ac}.
A general solution to its defining relation \eqref{defqr} is provided by \cite{Belov:2006jd}
\be
\qfD{\gamma} = \expe{-\frac12\,p^\Lambda \(q_\Lambda+A_{\Lambda\Sigma}p^\Sigma\)
+ \(q_\Lambda+A_{\Lambda\Sigma}p^\Sigma\) \theta_{\text{D}}^\Lambda
- p^\Lambda \phi_{{\text{D}},\Lambda}},
\label{quadraticrefinementpq}
\ee
where $\theta_{\text{D}}^\Lambda,\phi_{{\text{D}},\Lambda}$ are the so-called characteristics or generalized spin structure on $\CYm$,
defined modulo integers, and the terms proportional to the matrix $A_{\Lambda\Sigma}$ arise due to the change of the basis \eqref{symA}
and the non-integrality of charge $\gamma$.
Although one could think that the characteristics are just (half-integer) numbers, the symplectic invariance of
the D-instantons requires them to transform under symplectic rotations in order to keep $\qfD{\gamma}$ invariant,
\be
\label{sympchar}
Sp(2h_{1,1}+2,\IZ)\ni\rho={\scriptsize \begin{pmatrix} \cD & \cC \\ \cB & \cA \end{pmatrix}}\ :\quad
\begin{pmatrix} \theta_{\text{D}}^\Lambda \\ \phi_{{\text{D}},\Lambda} \end{pmatrix}
\ \mapsto\
\rho
\cdot \[
\begin{pmatrix} \theta_{\text{D}}^\Lambda \\ \phi_{{\text{D}},\Lambda} \end{pmatrix}
-\frac12
\begin{pmatrix} (\cA^T\cC)_d \\ (\cD^T\cB)_d  \end{pmatrix}
\],
\ee
where $(A)_d$ denotes the diagonal of a matrix $A$.

Now we are ready to present the discrete actions supposed to form the duality group of $\cM_H$.
Roughly, the idea is that one should take the parameters in the transformations \eqref{heis0}, \eqref{bjacr} and \eqref{SL2Z} to be integer.
Then they would correspond to large gauge transformations of the RR-gauge potentials and the B-field,
to monodromies around the large volume point, and to S-duality group of type IIB string theory, which are all expected to be symmetries
of the low-energy theory at full quantum level.
However, this naive idea requires some adjustments:
\begin{itemize}
\item
First, the correct form of the large gauge transformations is given by \cite{Bao:2010cc,Alexandrov:2010np}
\be
\opTH_{\eta^\Lambda,\tleta_\Lambda,\kappa}\ :\
\begin{array}{c}
\displaystyle{\zeta^\Lambda\ \mapsto\ \zeta^\Lambda+\eta^\Lambda,
\qquad
\tzeta_\Lambda\ \mapsto\ \tzeta_\Lambda+\tleta_\Lambda-A_{\Lambda\Sigma}\eta^\Sigma}
\\
\displaystyle{\sigma\ \mapsto\
\sigma + 2 \kappa- \tleta_\Lambda \(\zeta^\Lambda -2 \theta^\Lambda\)
+ \eta^\Lambda \(\tzeta_\Lambda +A_{\Lambda\Sigma}\zeta^\Sigma-2 \phi_\Lambda\)
- \eta^\Lambda \tleta_\Lambda. \rule{0pt}{17pt}}
\end{array}
\label{heisq}
\ee
Here $(\eta^\Lambda,\tleta_\Lambda,\kappa)\in \IZ^{2h_{1,1}+3}$, the $A$-dependent terms appear again as a consequence of \eqref{symA},
and  $\theta^\Lambda,\phi_\Lambda$ are the characteristics, similar to the ones appearing in \eqref{quadraticrefinementpq},
which characterize the fibration of the line bundle of the NS-axion over the torus of RR-scalars.

\item
Second, in \cite{Alexandrov:2010np} it was shown that the monodromies, given by the transformation \eqref{bjacr} with $\eps^a\in \IZ$,
should be accompanied by a shift of the NS-axion
\be
\sigma\ \mapsto\ \sigma + 2 \kappa (M_{\eps^a}),
\label{shiftM}
\ee
where $\kappa(M)$ is a character of the symplectic group. Since the monodromy subgroup is abelian, it can be represented as
$\kappa(M_{\eps^a}) = \kappa_a \epsilon^a $. The additional shift \eqref{shiftM} originates in the one-loop $g_s$-correction
which modifies the topology of the NS-axion line bundle over $\SK$.

\item
Finally, the S-duality group is represented by the transformations \eqref{SL2Z}
with $\gl{}\in SL(2,\IZ)$, which should be supplemented by a shift of the RR-scalar $\cla$ \cite{Alexandrov:2010ca}
\be
\cla\ \mapsto\ \cla \, - c_{2,a}\, \varepsilon(\gl{})\, ,
\label{cla}
\ee
where $\varepsilon(\gl{})$ is the logarithm of the multiplier system of the Dedekind eta function defined in appendix \ref{subap-Dedekind}.
This shift is closely related to the quantization conditions \eqref{fractionalshiftsD5} and is required to ensure that
the Heisenberg transformation with parameter $\eta^0$ coincides with the $SL(2,\IZ)$ transformation $\tau \mapsto \tau + \eta^0$.
\end{itemize}

\subsubsection{Corrected transformations and group law}
\label{subsec-correcttr}

It turns out that, even taking into account all the non-trivial adjustments described above,
the resulting set of discrete transformations is not satisfactory. As we show in appendix \ref{ap-Udual}, the generators
of these transformations do not really form a group (see \eqref{twistheisen})!\footnote{Of course, one could just {\it generate} a group
by taking products of all generators. But this would lead to a half-integer periodicity of RR-scalars
(in other words, one would have to allow $\tleta_\Lambda\in \hf\IZ$ in \eqref{heisq}),
which does not have any physical justification.}

The origin of this problem can be traced back to the characteristics appearing in the Heisenberg transformations \eqref{heisq}.
To see this, let us note that the monodromies \eqref{bjacr}, once we pass to the type IIA frame using \eqref{symA}, are
represented by the integer valued symplectic matrix
\be
\rho(M_{\eps^a})
=\(\begin{array}{cccc}
1\ & 0 & 0 & 0
\\
\epsilon^a & {\delta^a}_b & 0 & 0
\\
\ L_0(\epsilon) & L_b(\epsilon)+2A_{bc}\epsilon^c & 1 & \ -\epsilon^b\
\\
-L_a(\epsilon)\ & -\kappa_{abc}\epsilon^c & 0 & {\delta_a}^b
\end{array}\) ,
\label{monmat}
\ee
where we introduced two functions
\be
L_a(\eps)\equiv \frac12\, \kappa_{abc} \eps^b \eps^c-A_{ab}\eps^b,
\qquad
L_0(\eps)\equiv\frac16\, \kappa_{abc}\eps^a\eps^b\eps^c+\frac{1}{12}\, c_{2,a}\eps^a,
\ee
which are integer valued due to \eqref{propA} and \eqref{c2aprop}.
Since the characteristics $\theta^\Lambda,\phi_\Lambda$ should transform under symplectic rotations
as the D-instanton characteristics, they undergo a monodromy transformation which can be obtained
by plugging \eqref{monmat} into \eqref{sympchar}. Setting $\theta^\Lambda=0$, one eliminates some of the terms, but
even in this case one gets a non-trivial result
\be
\begin{split}
\phi_a \ \mapsto\ &\,\phi_a + \frac12\, \kappa_{aac} \epsilon^c,
\\
\phi_0 \ \mapsto\ &\,\phi_0 - \epsilon^a\phi_a
- \hf \(L_0(\epsilon) - \epsilon^a L_a(\epsilon) + \kappa_{aac} \epsilon^a  \epsilon^c\).
\end{split}
\label{transchar}
\ee
On the other hand, this is in contradiction with the fact that the monodromies can be obtained by commuting
$\eta^a$-Heisenberg shift with S-duality (see \eqref{maincom}) and that the characteristics are not expected
to transform under other isometries.

To resolve these inconsistencies, we note that the D-instanton characteristics can be absorbed into a redefinition
of the RR-fields and the NS-axion
\be
\begin{split}
\zeta^\Lambda-\thetaD^\Lambda\qquad\qquad \mapsto\ &\, \zeta^\Lambda,
\\
\tzeta_\Lambda-\phi_{{\text{D}},\Lambda}+A_{\Lambda\Sigma}\thetaD^\Sigma\qquad \mapsto\ &\, \tzeta_\Lambda,
\\
\sigma +\phi_{{\text{D}},\Lambda}\zeta^\Lambda-\thetaD^\Lambda\(\tzeta_\Lambda+A_{\Lambda\Sigma}\zeta^\Sigma\)\ \mapsto\ &\, \sigma.
\end{split}
\ee
This redefinition requires to modify the properties of these fields under symplectic transformations to take into account
the inhomogeneous terms in the corresponding transformations of characteristics \eqref{sympchar}.
In particular, this changes the monodromy transformations of $\tzeta_\Lambda$ and $\sigma$.
Instead of \eqref{bjacr} and \eqref{shiftM}, we can now take
\be
\label{bjacr-mod}
M_{\epsilon^a}\ :\quad  \begin{array}{l}
\displaystyle{b^a\ \mapsto\ b^a+\epsilon^a,
\qquad
\zeta^a\ \mapsto\ \zeta^a + \epsilon^a \zeta^0,}
\\
\displaystyle{\tzeta_a\ \mapsto\ \tzeta_a -\kappa_{abc}\zeta^b \epsilon^c
-\frac12\,\kappa_{abc} \epsilon^b \epsilon^c \zeta^0+A_{ab}\eps^b ,}
\\
\displaystyle{\tzeta_0\ \mapsto\ \tzeta_0 -\tzeta_a \epsilon^a+\frac12\, \kappa_{abc}\zeta^a \epsilon^b \epsilon^c
+\frac16\,\kappa_{abc} \epsilon^a \epsilon^b \epsilon^c \zeta^0
-\hf\, A_{ab}\epsilon^a\epsilon^b+\frac{c_{2,a}}{8}\, \epsilon^a,}
\\
\displaystyle{\,\sigma\ \mapsto\ \sigma -A_{ab}\eps^a\zeta^b-\hf\( A_{ab}\eps^a\eps^b+\frac14\, c_{2,a}\eps^a\)\zeta^0 +2\kappa_a \epsilon^a.}
\end{array}
\ee
A new input, which leads to an improvement of the duality group representation, is that
we require that the {\it new} redefined fields are related to the type IIB coordinates, transforming under S-duality according to
\eqref{SL2Z} and \eqref{cla}, by the {\it standard} classical mirror map \eqref{symptobd}.
Thus, we change transformations of some fields ($\cla$ and $\cl0$) under monodromies and leave other transformations
unmodified. All characteristics can now be set to zero.\footnote{More precisely, one can still have non-vanishing characteristics
$\theta^\Lambda,\phi_\Lambda$ which transform now homogeneously under monodromies. However, one can check that
the group law fixes them to zero. Non-vanishing values can appear only if one relaxes \eqref{valA}. For instance, one has $\phi_0=\hf\, A_{00}$.}

\begin{table}
\vspace{0.cm}\hspace{-1.cm}
\begin{tabular}{|c|c|c|c|c|c|}
\hline $\vphantom{\frac{A^{A^A}}{A_{A_A}}}$
& $b^a$ & $ c^a$ & $\cla$ &$\cl0$ & $\psi$
\\
\hline $\vphantom{\frac{A^{A^A}}{A_{A_A}}}$
$S$ &  $c^a$     &  $-b^a$  & $\cla+\frac{c_{2,a}}{8}$
& $-\psi$ & $\cl0$
\\
\hline $\vphantom{\frac{A^{A^A}}{A_{A_A}}}$
$T$ &  $b^a$     &  $c^a+b^a$  & $\cla-\frac{c_{2,a}}{24}$
& $\cl0$ & $\psi-\cl0$
\\
\hline $\vphantom{\frac{A^{A^A}}{A_{A_A}}}$
$\opT^{(1)}_{\epsilon^a,0}$ &   $b^a+\epsilon^a$    &   $c^a$  & $\cla+\frac12\, \kappa_{abc} \epsilon^b c^c+ A_{ab} \epsilon^b$
& $\begin{array}{c}\cl0-\epsilon^a \cla \\-\frac16\, \kappa_{abc} \epsilon^a (b^b+2\epsilon^b)c^c \\
- \frac12 A_{ab}\epsilon^a \epsilon^b +\frac{c_{2,a}}{8}\, \epsilon^a\end{array}$
& $\psi+\frac16\, \kappa_{abc}\epsilon^a c^b c^c- \kappa_a \epsilon^a$
\\
\hline $\vphantom{\frac{A^{A^A}}{A_{A_A}}}$
$\opT^{(1)}_{0,\eta^a}$&  $b^a$    &    $c^a+\eta^a$  & $\cla-\frac12\, \kappa_{abc} \eta^b b^c+A_{ab} \eta^b$
&  $\cl0+\frac16\, \kappa_{abc} \eta^a b^b b^c + \frac{c_{2,a}}{24}\, \eta^a $
&  $\begin{array}{c}\psi+\eta^a \cla+\frac12\, A_{ab} \eta^a \eta^b\\
-\frac16\, \kappa_{abc}\eta^a b^b (c^c+2\eta^c)\end{array}$
\\
\hline $\vphantom{\frac{A^{A^A}}{A_{A_A}}}$
$\opT^{(2)}_{\tilde\eta_a}$
&  $b^a$  & $c^a$ & $\cla+\tilde\eta_a$
& $\cl0$  & $ \psi$
\\
\hline $\vphantom{\frac{A^{A^A}}{A_{A_A}}}$
$\opT^{(3)}_{\tilde\eta_0,\kappa}$ &  $b^a$  & $c^a$ & $\cla$
&  $\cl0+\tilde\eta_0$ & $\psi+\kappa$
\\
\hline
\end{tabular}
\caption{The action of generators of the discrete symmetry transformations in the type IIB coordinate basis.}
\label{tab-U}
\end{table}
\vspace{0.cm}

We summarize the resulting action of all generators of the duality group in the type IIB coordinate basis in Table \ref{tab-U}.
This table should be supplemented by the standard $SL(2,\IZ)$ action \eqref{SL2Z} on the variables $\tau$ and $t^a$
which are not affected by other transformations.
It uses the following notations for generators:
$\opT^{(0)}_{\eta^0}$, $\opT^{(1)}_{0,-\eta^a}$, $\opT^{(2)}_{\tilde\eta_a}$ and $\opT^{(3)}_{\tilde\eta_0,-\kappa}$
correspond to the generators of the Heisenberg subgroup $\opTH_{\eta^\Lambda,\tleta_\Lambda,\kappa}$,
$\opT^{(1)}_{\epsilon^a,0}=M_{\eps^a}$ is the monodromy generator, and $SL(2,\IZ)$ is generated by
\be
S=\(\begin{array}{cc}
0 & -1
\\
1 & 0
\end{array}\),
\qquad
T=\(\begin{array}{cc}
1 & 1
\\
0 & 1
\end{array}\).
\label{genST}
\ee
However, the action of the Heisenberg shift $\opT^{(0)}_{1}$ is identical to $T$ and therefore it is not presented in the table.
These notations indicate that the generators $\opT^{(n)}$ with $n>0$ form a graded nilpotent subgroup where each $n$th level
forms a representation of $SL(2,\IZ)$. This fact has a direct relation to the split of non-perturbative corrections
into S-duality invariant sectors presented in \eqref{quantcor}.

In appendix \ref{ap-Udual} we demonstrate that the transformations given in the above table
satisfy the group law provided one fixes the character of the monodromy group as
\be
\kappa_a=-\frac{c_{2,a}}{24}\, .
\label{indentkappa}
\ee
One of the most important group relations is given by
\be
S^{-1}\, \opT^{(1)}_{0,\eta^a}\, S=\opT^{(1)}_{\eta^a,0}.
\label{maincom}
\ee
This relation ensures that the fivebrane instantons generated via S-duality are guaranteed to be compatible with other isometries.
And indeed, in appendix \ref{ap-MHinv} we will prove that the transformations found in this section, unlike
the previous ones, are consistent with our results for fivebrane instanton corrections.

\section{QK manifolds in the twistor approach}
\label{sec-twistor}

\subsection{Twistorial construction of QK manifolds}

To incorporate instanton corrections to the geometry of the HM moduli space consistently with its QK property,
it is instrumental to use the twistorial construction of such manifolds \cite{MR664330,MR1327157,Alexandrov:2008nk}.
As we review below, it allows to encode any QK metric in a set of holomorphic data on the twistor space $\cZ$,
which is constructed as a canonical $\CP$ bundle over the original manifold $\cM$.
Whereas $\cM$ carries a triplet of non-integrable almost complex structures, $\cZ$ is a K\"ahler manifold.
Furthermore, it has a {\it complex contact structure}
defined globally by the kernel of the following $(1,0)$ form
\bea
D t = \text{d} t + p_+ - \I p_3 t + p_- t^2 ,
\eea
where $t$ is the fiber coordinate on $\IC\IP^1$ and $(p_{\pm},p_3)$
are the $SU(2)$ part of Levi-Civita connection on $\cM$.
It is more convenient however to use a local description of this structure in which case it can be represented by
a holomorphic one-form $\cX^{[i]}$ having the same kernel as $Dt$.
Here the upper index shows that this one-form is defined only in a patch $\cU_i$ of
an atlas covering the twistor space, $\cZ=\cup\cU_i$.

The contact form $\cX^{[i]}$ allows to define a set of local Darboux coordinates such that
\be
\cX^{[i]} = \text{d} \ai{i} + \xii{i}^\Lambda \text{d} \txii{i}_\Lambda .
\ee
Then the contact structure, and the full geometry of $\cM$, is completely determined
by the {\it contact transformations} relating the Darboux coordinate systems
on the overlaps of two patches $\cU_i\cap\cU_j$ and preserving the contact one-form up to a non-vanishing holomorphic factor.
One way to parametrize such transformations is to use holomorphic functions $\Hij{ij}(\xii{i},\txii{j},\ai{j})$
which depend on $\xi^\Lambda$ in patch $\cU_i$ and $\txi_\Lambda,\alpha$ in patch $\cU_j$.
Then the gluing conditions between Darboux coordinates read as follows \cite{Alexandrov:2009zh}
\be
\begin{split}
\xii{j}^\Lambda = &\,  \xii{i}^\Lambda -\p_{\txii{j}_\Lambda }\Hij{ij}
+\xii{j}^\Lambda \, \p_{\ai{j} }\Hij{ij} ,
\\
\txii{j}_\Lambda =&\,  \txii{i}_\Lambda
 + \p_{\xii{i}^\Lambda } \Hij{ij} ,
\\
\ai{j} = &\, \ai{i}
 + \Hij{ij}- \xii{i}^\Lambda \p_{\xii{i}^\Lambda}\Hij{ij} ,
\end{split}
\label{QKgluing}
\ee
and result in the following transformation of the contact one-form
\be
\label{glue2}
\CX\ui{j} =  \(1-\p_{\ai{j} }\Hij{ij}\)^{-1}  \CX\ui{i}.
\ee
Supplementing \eqref{QKgluing} by appropriate reality and regularity conditions,
these discrete equations can be rewritten as a system of integral equations which relate the Darboux coordinates
to the integrals along contours on $\CP$ of the discontinuities from \eqref{QKgluing} multiplied by a certain $t$-dependent kernel.
Their solution provides the Darboux coordinates as functions of the fiber coordinate $t$ and coordinates
on the base $\cM$ of the twistor fibration.
Then a straightforward but tedious procedure leads to the QK metric on $\cM$ \cite{Alexandrov:2008nk}.

Thus, the QK geometry turns out to be encoded in a set of holomorphic functions $\Hij{ij}$, which we call {\it transition functions},
and the associated set of contours on $\CP$. Typically, the contours separate the two patches whose Darboux coordinates are related by the contact
transformation generated by $\Hij{ij}$.
It is important to note that in this construction both closed and open contours may appear, as is the case, for instance,
in the twistorial description of the HM moduli space.

\subsection{Contact bracket}
\label{subsec-contact}

The twistorial construction presented above relies on the parametrization of contact transformations in terms of transition functions $\Hij{ij}$.
Although such parametrization is very explicit, the main obstacle in dealing with it comes from the fact that
the arguments of $\Hij{ij}$ belong to different patches. As a result, even simple-looking gluing conditions may be generated by
complicated transition functions. This issue becomes particularly problematic when one tries to describe the action
of some symmetries on the twistor data.
Typically such an action is most naturally formulated in terms of Darboux coordinates in one patch,
and it can become highly non-linear being written as a symmetry transformation of $\Hij{ij}$.
Below we will see several examples of such situation.

This complication can be avoided if one uses an alternative parametrization which we proposed in \cite{Alexandrov:2014mfa}.
It is based on the so-called {\it contact bracket} which is an extension of the Poisson bracket construction to the domain of contact geometry.
The contact bracket maps two local sections $\muh_1\in\cO(2m)$ and $\muh_2\in\cO(2n)$ to a local
section of $\cO(2(m+n-1))$ line bundle, given in terms of Darboux coordinates by \cite{Alexandrov:2008gh}
\be
\label{poisson}
\begin{split}
\{ \muh_1, \muh_2 \}_{m,n}= &\,
\pa_{\xi^\Lambda}  \muh_1 \pa_{\txi_\Lambda}  \muh_2 +
\(m \muh_1  -\xi^\Lambda \pa_{\xi^\Lambda} \muh_1\)\pa_\alpha\muh_2
\\
&\, - \pa_{\xi^\Lambda}  \muh_2 \pa_{\txi_\Lambda}  \muh_1
-\(n \muh_2-\xi^\Lambda \pa_{\xi^\Lambda} \muh_2\) \pa_\alpha \muh_1 .
\end{split}
\ee
It is easy to check that this bracket satisfies the standard Jacobi identity, skew-symmetry and Leibnitz rule provided one keeps track
of the geometric nature of all objects. For instance, the Leibnitz rule for $\mu_1,\mu_2$ defined as above and $\mu_3\in \cO(2k)$ reads as
\be
\{\mu_1 \mu_2 , \mu_3 \}_{m+n,k}
= \mu_1 \{\mu_2,\mu_3\}_{n,k} + \mu_2 \{\mu_1,\mu_3 \}_{m,k} .
\ee
We mostly need the specialization of \eqref{poisson} to the case $(m,n)=(1,0)$ which provides the action of
a vector field $\Xf{\mu_1}$ with the (generalized) moment map $\mu_1$ on a local complex function $\mu_2$ \cite{MR872143}.
Setting $\mu_1=\hH$ and $\mu_2$ to be one of the Darboux coordinates, one explicitly finds\footnote{If it is not indicated explicitly,
in the following the bracket $\{\, \cdot\, ,\, \cdot \}$ will always mean the contact bracket between $\cO(2)$
and $\cO(0)$ sections, i.e. of type (1,0).}
\be
\begin{split}
\{\hH,\xi^\Lambda\}=&\, -\p_{\txi_\Lambda} \hH+\xi^\Lambda\p_\alpha \hH,
\qquad
\{\hH,\txi_\Lambda\}=\p_{\xi^\Lambda} \hH,
\\
&\qquad
\{\hH,\alpha\}=\hH-\xi^\Lambda\p_{\xi^\Lambda} \hH.
\end{split}
\label{contbr}
\ee
Note that in the case where this bracket is evaluated on sections (of different bundles) represented by the same function,
despite the skew-symmetry property, the result is non-vanishing and is given by
\be
\{h,h\} =h\p_\alpha h.
\ee
Another important property, which plays a crucial role in our construction, is the behavior of \eqref{contbr}
under contact transformations. If $\vrh$ is such transformation mapping $\cX\mapsto \lambda\cX$ then
\be
\vrh\cdot \{h,f\}=\{ \lambda^{-1}\vrh\cdot h, \vrh\cdot f\}.
\label{trans-contact}
\ee
This property generalizes the familiar invariance of the Poisson bracket under canonical transformations
to the realm of contact geometry.
We provide its proof in appendix \ref{ap-contactbr} in a coordinate independent way.

The importance of the contact bracket becomes clear if one considers the action
of the vector field $\Xf{\hH}=\{\hH,\,\cdot\,\}$ on the contact one-form,
which is found to be
\be
\cL_{\Xf{\hH}} \cX
= (\p_\alpha \hH) \cX.
\label{tr-cb-X}
\ee
This means that it generates an infinitesimal contact transformation.
Furthermore, identifying $h$ with vanishingly small transition functions $\Hij{ij}$, one observes that
\eqref{contbr} and \eqref{tr-cb-X} represent a linearized version of \eqref{QKgluing} and \eqref{glue2}, respectively.
Therefore, {\it any} infinitesimal contact transformation can be generated in this way
and a finite transformation can be obtained by exponentiation.
Thus, we can rewrite the gluing conditions \eqref{QKgluing}
as
\be
\Xi^{[j]} = \exp\({\Xf{\hHij{ij}}}\) \cdot \Xi^{[i]} ,
\label{newglue}
\ee
where $\Xi^{[i]}$ denotes the set of Darboux coordinates in patch $\cU_i$.
This formula provides a parametrization of contact transformations in terms of functions
$\hHij{ij}$, which we call {\it contact Hamiltonians}\footnote{Note that we changed a bit the terminology as in \cite{Alexandrov:2014mfa}
we called $\hHij{ij}$ ``improved transition functions".}
and which, in contrast to the ordinary transition functions, are considered as functions of coordinates in one patch only.
As we will see below, this parametrization crucially simplifies
various properties and results.

A relation between $\hHij{ij}$ and $\Hij{ij}$ can be found by comparing the gluing conditions \eqref{newglue} and \eqref{QKgluing}.
Recombining some of these equations, one can get an explicit formula for transition functions in terms of the action
generated by contact Hamiltonians on the Darboux coordinates
\be
\Hij{ij}=\(e^{\Xf{\hHij{ij}}}-1\)\ai{i}+\xii{i}^\Lambda\( e^{\Xf{\hHij{ij}}}-1\)\txii{i}_\Lambda.
\label{relHH}
\ee
Note however that this expression computes $\Hij{ij}$ as a function of Darboux coordinates in patch $\cU_i$, whereas we need
to transfer $\txi_\Lambda$ and $\alpha$ to patch $\cU_j$ to be able to compute the derivatives entering the gluing conditions \eqref{QKgluing}.
Therefore, it is indispensable to compute the full contact transformation and not only the combination \eqref{relHH}.
In the particular case of $\hHij{ij}$ independent of $\txi_\Lambda$ and $\alpha$, the two objects coincide, $\Hij{ij}=\hHij{ij}(\xi)$,
and this problem does not arise.

\subsection{Gauge transformations}
\label{subsec-gauge}

A fact which will play an important role below is that the contact structure does not fix the Darboux coordinates uniquely,
but has a freedom to perform {\it local} contact transformations.
Such a ``gauge" transformation affects not only the Darboux coordinates, but also the transition functions
and the corresponding contact Hamiltonians. Here we want to display this action.

As any contactomorphism, in each patch the gauge transformation can be parametrized by a holomorphic function
in one of the two ways we described above:
either as in \eqref{QKgluing} or via the contact bracket as in \eqref{newglue}.
Let us choose the second way and denote the corresponding holomorphic functions by $\gi{i}$.
A crucial difference with the contact Hamiltonians is that $\gi{i}$ must be regular in $\cU_i$ in order
to preserve the regularity of the Darboux coordinates.
The contact Hamiltonian in the gauge transformed picture, $\hHij{ij}_g$, satisfies
\be
\exp\(\Xf{\hHij{ij}_g}\)
= e^{-\Xf{\gi{i}} } \, \exp\Bigl(\Xf{\hHij{ij}} \Bigr)\,e^{\Xf{\gi{j}} } .
\label{gaugetr}
\ee
Although it can in principle be extracted using the Baker-Campbell-Hausdorff formula, the result does not appear to be explicit.
In fact, in this paper we will need only a particular case of \eqref{gaugetr} where the gauge transformation functions
are the same in all patches, $\gi{i}=g$.
Then applying
\be
\[\Xf{g},\Xf{h}\]=\Xf{\{g,h\}_{1,1}^{}},
\label{commXXh}
\ee
which is nothing else but the Jacobi identity for the contact bracket,
the contact Hamiltonian $\hHij{ij}_g$ can be computed explicitly and is given by
\be
\hHij{ij}_{g}=e^{-\{ g,\,\, \cdot\,\, \}_{1,1}^{}}\cdot \hHij{ij}.
\ee
Furthermore, if $g$ depends only on $\xi^\Lambda$, the effect of the gauge transformation is just the shift of the arguments
of the contact Hamiltonian
\be
\hHij{ij}_g=\hHij{ij}\(\xi^\Lambda\, ,\, \txi_\Lambda - \p_{\xi^\Lambda} g \, ,\, \alpha-g+\xi^\Lambda\p_{\xi^\Lambda} g \).
\label{gaugehHxi}
\ee
The corresponding formula for the gauge transformed transition function $\Hij{ij}_g$ can be obtained either via \eqref{relHH}
or directly by applying the gauge transformation to the gluing conditions \eqref{QKgluing}. Both ways lead to the same result,
but since it is a bit complicated and not needed for our purposes, we refrain from giving it here.

\subsection{S-duality in twistor space}
\label{subsec-Sdual}

Finally, we discuss the constraints on the twistor data imposed by the presence of
the $SL(2,\IZ)$ isometry group on the QK manifold $\cM$.
We assume that there are coordinates in which the $SL(2,\IZ)$ action is given as in \eqref{SL2Z} and \eqref{cla}.

It is known that any isometry on $\cM$ can be lifted to a {\it holomorphic} action on the twistor space.
The lift of $SL(2,\IZ)$, without assuming that $\cM$ has any additional continuous isometries,
has been obtained in \cite{Alexandrov:2013mha}
and is provided by the following transformation of the fiber coordinate
\be
\label{SL2varpi}
\varpi\ \mapsto\  \gl{}\[\htm^{-c,a}\]\frac{t-\htp^{c,d}}{t-\htm^{c,d}}\, ,
\ee
where $\htpm^{c,d}$ are the two roots of the equation $c\xi^0 (t) +d=0$.
Then the resulting $SL(2,\IZ)$ action is isometric if the Darboux coordinates transform as follows \cite{Alexandrov:2008gh}
\be
\label{SL2Zxi}
\begin{split}
&\xi^0\mapsto \frac{a \xi^0 +b}{c \xi^0 + d} \, ,
\qquad
\xi^a \mapsto \frac{\xi^a}{c\xi^0+d} \, ,
\qquad
\txi_a \mapsto \txi_a +  \frac{c}{2(c \xi^0+d)} \kappa_{abc} \xi^b \xi^c- c_{2,a}\, \varepsilon(\gl{})\, ,
\\
&\begin{pmatrix} \txi_0 \\ \alpha \end{pmatrix}\mapsto
\begin{pmatrix} d & -c \\ -b & a  \end{pmatrix}
\begin{pmatrix} \txi_0 \\ \alpha \end{pmatrix}
+ \frac{1}{6}\, \kappa_{abc} \xi^a\xi^b\xi^c
\begin{pmatrix}
c^2/(c \xi^0+d)\\
-[ c^2 (a\xi^0 + b)+2 c] / (c \xi^0+d)^2
\end{pmatrix} .
\end{split}
\ee
Indeed, such transformation ensures that the contact one-form is only rescaled by a holomorphic factor
\be
\cX\ \mapsto\ \frac{\cX}{c \xi^0 +d}.
\label{Strans-X}
\ee
Thus, it represents an example of a holomorphic contact transformation and,
since it preserves the contact structure, it also preserves the metric.

The question we are interested in is: which twistor data, namely the contours and transition functions,
ensure the transformations \eqref{SL2Zxi}?
In \cite{Alexandrov:2012bu,Alexandrov:2013mha} it was shown that \eqref{SL2Zxi} holds if the twistor data
can be split into two parts. The first part gives a ``classical" space which is in fact identical to
$\cM_H$ in the classical, large volume limit. It is defined by the two transition functions
\be
\Hij{+0} =\Fcl(\xii{+}),
\qquad
\Hij{-0}=\bFcl(\xii{-}),
\label{treeHij}
\ee
where $\Fcl(X)=-\kappa_{abc}\,\frac{X^aX^bX^c}{6X^0}$ is the classical part of the holomorphic prepotential \eqref{lve},
associated with the contours around the north ($t=0$) and south ($t=\infty$) poles of $\CP$, respectively.
The second part can be viewed as ``quantum corrections" to $\cM_H$ and consists of
the contours $C_{m,n;i}$ and the corresponding transition functions $\Hij{i}_{m,n}$,
labeled by a pair of integers $(m,n)$ and additional index $i$. To preserve $SL(2,\IZ)$,
they should be such that $C_{m,n;i}$ are mapped into each other as
\be
C_{m,n;i} \ \mapsto\ C_{\hat m,\hat n;i},
\qquad
\( \hat m\atop \hat n\) =
\(
\begin{array}{cc}
d & -c
\\
-b & a
\end{array}
\)
\( m \atop n \) ,
\label{mappatches}
\ee
whereas $\Hij{i}_{m,n}$ satisfy a {\it non-linear} transformation property given
explicitly in appendix \ref{ap_Sdualconstrap} (see \eqref{Sdualconstr}).
However, the same constraint considerably simplifies once it is rewritten in terms of the contact Hamiltonians $\hHij{i}_{m,n}$,
consistently with the expectations of section \ref{subsec-contact}.
Indeed, since \eqref{SL2Zxi} is a contact transformation, one can apply the property \eqref{trans-contact}
of the contact bracket where $\lambda=(c\xi^0+d)^{-1}$ due to \eqref{Strans-X}.
As a result, it turns out that, to generate the Darboux coordinates satisfying \eqref{SL2Zxi},
the contact Hamiltonians should follow a simple {\it linear} transformation
\cite{Alexandrov:2014mfa}\footnote{If $C_{m,n;i}$ are closed contours, it is possible also that the result of the transformation
has in addition some regular contributions, which can then be absorbed
by a gauge transformation described in section \ref{subsec-gauge}
into a redefinition of Darboux coordinates not affecting the contact structure.\label{foot-regular}}
\be
\hHij{i}_{m,n}\ \mapsto\ \frac{\hHij{i}_{m',n'}}{c\xi^0+d},
\qquad
\begin{pmatrix} m'\\ n' \end{pmatrix} =
\begin{pmatrix}
a & c
\\
b & d
\end{pmatrix}
\begin{pmatrix} m \\ n \end{pmatrix}.
\label{transhH}
\ee

This provides an explicit example how the contact bracket formalism simplifies various aspects of the twistorial description
of QK manifolds.
Furthermore, since any isometry is realized on the twistor space as a contact transformation, the property \eqref{trans-contact}
ensures that the passage to the contact Hamiltonians linearizes any symmetry action.

\section{D-instantons in twistor space}
\label{sec-Dinst}

\subsection{D-instantons in type IIA picture}

The D-instanton corrections to the HM metric can be incorporated using the twistor framework
presented in the previous section. The most elegant formulation they obtain
in the type IIA picture \cite{Alexandrov:2008gh,Alexandrov:2009zh}
where they are induced by D2-branes wrapping special Lagrangian submanifolds of $\CYm$.

First, we note that the twistor description of the tree level metric on $\cM_H$ can be obtained starting
from the transition functions \eqref{treeHij} where $\Fcl$ should be replaced by the full prepotential \eqref{lve}.
To incorporate contributions from the D-instantons, we introduce the contours on $\CP$ known as {\it BPS rays},
which extend from the north to the south pole along the direction determined by the central charge $Z_\gamma$ \eqref{defZ}
\be
\ell_\gamma=\{\varpi:\ Z_\gamma(z)/t\in\I\IR^-\}.
\label{BPSray}
\ee
With these contours we associate the contact Hamiltonians
\be
\hHij{\gamma}(\xi,\txi)=H_{\gamma}(\Xi_\gamma),
\qquad
H_{\gamma}(\Xi_{\gamma})
=  \frac{\bar \Omega(\gamma)}{4 \pi^2}\, \sigma_D (\gamma)  \expe {-\Xi_{\gamma}},
\label{Hgam}
\ee
where $\Xi_\gamma=q_\Lambda \xi^\Lambda-p^\Lambda\txi_\Lambda$,
the coefficients $\bar\Omega(\gamma)$ are the so-called rational Donaldson-Thomas invariants \cite{Manschot:2010xp,Manschot:2010qz}
\be
\bar\Omega(\gamma) = \sum_{d\vert\gamma} \frac{1}{d^2} \, \Omega(\gamma/d),
\label{rationalinvariants}
\ee
and $\sigma_D(\gamma)$ is the quadratic refinement \eqref{quadraticrefinementpq} with all characteristics set to zero
(see section  \ref{subsec-correcttr}).
These contact Hamiltonians, via \eqref{newglue}, generate contact transformations between Darboux coordinates on the two sides of the BPS rays,
thereby changing the contact structure and deforming the metric so that the leading corrections take the expected form \eqref{d2quali}.

Note that the operators $e^{\Xf{\hHij{\gamma}}}$ generating the contact transformations induced by \eqref{Hgam}
are nothing else but a lift to the contact geometry
of the Kontsevich-Soibelman (KS) operators\footnote{More precisely, the usual KS operators are obtained if in \eqref{Hgam}
the rational DT invariants are replaced by the usual ones and the exponential is replaced by the dilogarithm.
However, the product over all (collinear) charges, which enters the wall-crossing formula, is the same in the two versions.}
$U_\gamma^{\bar\Omega(\gamma)}$
satisfying the wall crossing formula \cite{ks}. It dictates how the DT invariants
change after crossing a wall of marginal stability in the special \kahler moduli space of $z^a$
and ensures the smoothness of the moduli space metric across the walls \cite{Gaiotto:2008cd}.
Provided $\Gamma(z)$ is a set of charges for which $Z_\gamma(z)$ become aligned at point $z^a$ and
$\bar\Omega^\pm(\gamma)$ are the rational DT invariants on the two sides of the wall,
the KS formula states that
\be
\label{ewall}
\prod^\ccwarrow_{\gamma\in \Gamma(z)}
U_{\gamma}^{\bar\Omega^-(\gamma)}
=
\prod^\cwarrow_{\gamma\in \Gamma(z)}
U_{\gamma}^{\bar\Omega^+(\gamma)},
\ee
where the two products are taken in the opposite order. (In both cases the order corresponds to decreasing the phase of $Z_\gamma$
at a given point in the moduli space.)
The fact that this formula extends from the operators generating symplectomorphisms to the level of contact transformations
was proven in \cite{Alexandrov:2011ac} using dilogarithm identities, which in turn follow from the classical limit of the motivic
version of \eqref{ewall}.

Another comment is that one can easily compute the transition function corresponding to the contact Hamiltonian \eqref{Hgam}.
Using \eqref{relHH} and the properties of the contact bracket \eqref{contbr}, one finds
\be
\Hij{\gamma} = H_{\gamma} - \frac12\, q_{\Lambda}p^\Lambda  \(H'_{\gamma}\)^2,
\label{trHij}
\ee
where the prime means the derivative.
This is the form in which the D-instanton corrections have been first formulated to all orders in the instanton expansion
in \cite{Alexandrov:2009zh}. Note again the simplicity and symplectic invariance of the contact Hamiltonians
in contrast to (the absence of) the corresponding properties of the transition functions.

\subsection{D1-D(-1)-instantons and S-duality}

Although the formulation presented in the previous subsection is very simple and incorporates all D-instanton effects, it is not
suitable for our purposes. In the next section we are going to apply S-duality to derive fivebrane instantons. Therefore, we need
a formulation which respects this symmetry, whereas the above type IIA picture is rather adapted to symplectic invariance.

\lfig{Example of the BPS rays of D5 (red), D1 (green) and D(-1) (brown) branes and the effect of the gauge transformation
which rotates the two latter types of rays to the real line. $\cU_\gamma$ denotes the patch lying in the counterclockwise direction from
the BPS ray $\ell_\gamma$.}{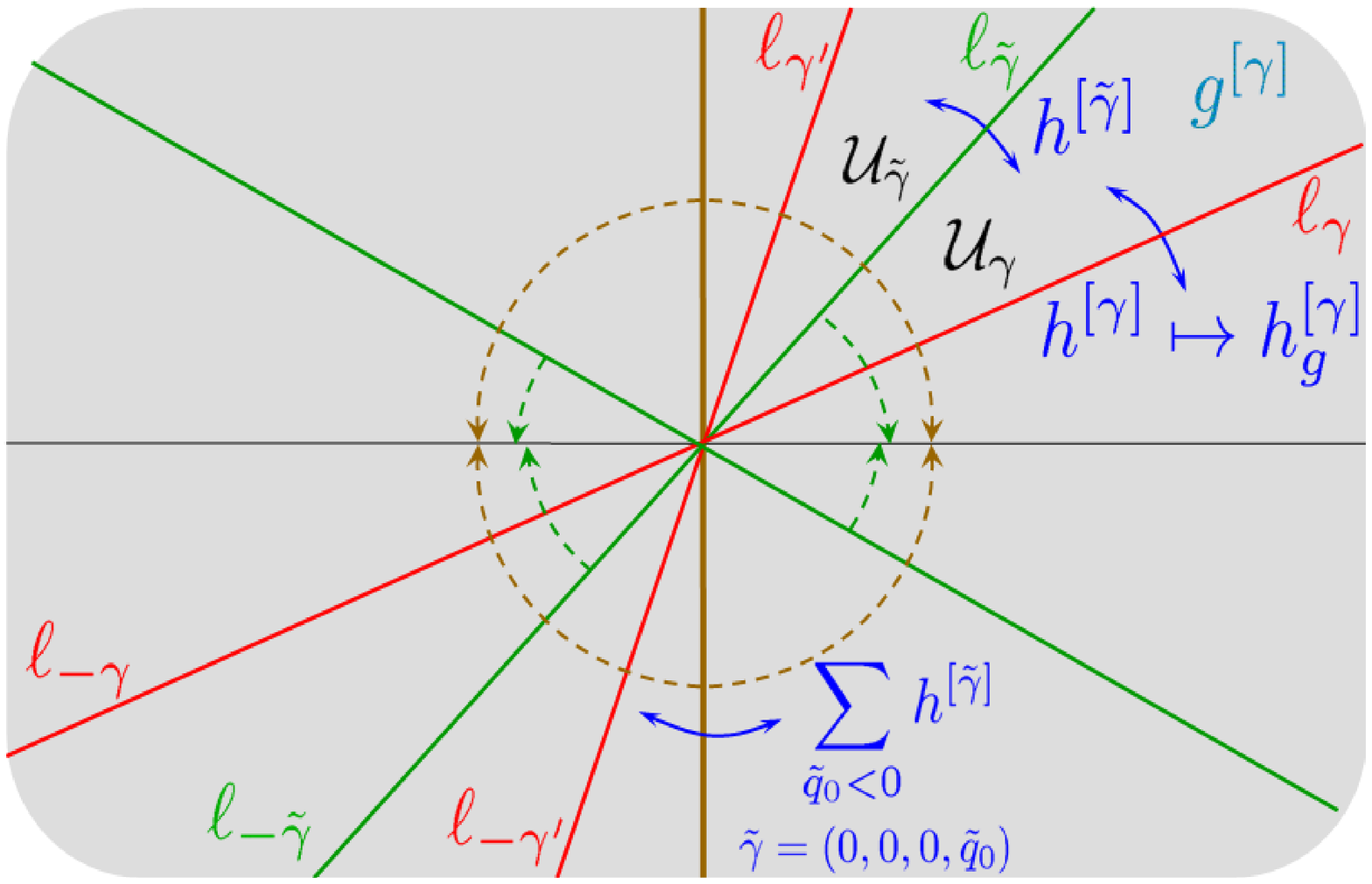}{12cm}{fig-Dinst-gauge}{-0.6cm}

For the two sectors corresponding to D1 and D(-1)-instantons (see \eqref{quantcor}), the passage to a manifestly S-duality invariant formulation
was understood in \cite{Alexandrov:2009qq,Alexandrov:2012bu}. The idea is to perform a gauge transformation on the twistor space
such that the gauge transformed twistor data satisfy the constraints spelled in section \ref{subsec-Sdual}, which ensure the presence
of the $SL(2,\IZ)$ isometry.
To display the corresponding gauge transformation, we need to introduce some definitions.
First, let us define an ordering on the charge lattice according to the phase of the central charge function
saying that $\gamma>\gamma'$ if $\pi>\arg\(Z_\gamma Z_{\gamma'}^{-1}\)>0$.
Then for each charge $\gamma$ we define an associated set of D(-1)-brane charges
whose BPS rays lie in the same half-plane as $\ell_\gamma$
\be
\GamD{-1}_\gamma=\left\{\gamD{-1}=(0,0,0,\tilde q_0)\ :\
\tilde q_0\Re Z_\gamma>0
\right\},
\label{latDm1}
\ee
and another set of D1-brane charges
for which the BPS rays are between $\ell_\gamma$ and the imaginary axis
\be
\GamD{1}_\gamma=\left\{\gamD{1}=(0,0,\tilde q_a,\tilde q_0)\in H_2^+\cup H_2^-\ :\
\Nq(\gamD{1})=\Nq(\gamma) \ {\rm and}\
\begin{array}{c}
\gamD{1}> \gamma \quad \mbox{for}\ \Nq(\gamma)\ {\rm odd}
\\
\gamD{1}\le \gamma \quad \mbox{for}\ \Nq(\gamma)\ {\rm even}
\end{array}
\right\},
\label{latD1}
\ee
where $H_2^+$ is the set of charges corresponding to effective homology classes on $\CYm$,
$H_2^-$ is the set of opposite charges, and $\Nq(\gamma)$ denotes the quadrant which
$\ell_\gamma$ belongs to.\footnote{One can write
$\Nq(\gamma)=\left\lfloor\frac{2}{\pi}\, \arg \(\I Z_\gamma\)\right\rfloor$.}
Note that both the ordering and the two charge sets $\GamD{\pm 1}_\gamma$ may change after crossing a wall of marginal stability.
Given these definitions, we define a holomorphic function which generates the gauge transformation in the patch
$\cU_\gamma$ taken to lie in the counterclockwise direction from the BPS ray $\ell_\gamma$ (see Fig. \ref{fig-Dinst-gauge})
\be
\gi{\gamma}=(-1)^{\Nq(\gamma)}\[\frac{1}{2}\sum_{\gamD{-1}\in\GamD{-1}_\gamma} \hHij{\gamD{-1}}+
\sum_{\gamD{1}\in\GamD{1}_\gamma}\hHij{\gamD{1}}\].
\label{fungengauge}
\ee
This gauge transformation has a very simple geometric meaning: It simply
rotates the BPS rays corresponding to D1-instantons either to the positive or negative real axis
depending on which one is the closest to the given ray. On the other hand, the D(-1) BPS rays,
which all go along the imaginary axis, are split into two ``halves" which are also rotated to the two real half-axes.

As a result, the contours associated with all D1 and D(-1)-branes coincide with either positive or negative real axis
and the corresponding contact Hamiltonians or transition functions (which are the same in this case since they depend only on $\xi^\Lambda$)
can be summed up. Furthermore, a Poisson resummation of this series over $\tilde q_0$ provides an alternative twistor description
fitting the constraints of S-duality \cite{Alexandrov:2009qq,Alexandrov:2012bu}. Instead of BPS rays, we can now consider
the contours $C_{m,n}$ centered around the points $\htp^{m,n}$ defined below \eqref{SL2varpi},
whereas the corresponding contact Hamiltonians are given by
\be
\hH_{m,n}^{\rm D1}(\xi)
= -\frac{\I}{(2\pi)^3}\!\!\sum_{q_a\in H_2^+\cup\{0\}}\!\!\!\! n_{q_a}^{(0)}\,
\begin{cases}
\displaystyle\frac{e^{-2\pi \I m q_a\xi^a}}{m^2(m\xi^0+n)}, &
\quad  m\ne 0,
\\
\displaystyle (\xi^0)^2 \,\frac{e^{2\pi \I n q_a\xi^a/\xi^0}}{n^3}, &\quad  m=0.
\end{cases}
\label{defG1}
\ee
Here we set  $\gfinv_{0}=-\chi_\CYm/2$ and used that
\be\label{instnr}
\begin{split}
\Omega(\gamD{1}) =&\, \gfinv_{q_a} \ \quad\mbox{\rm for}\quad \gamD{1}=(0,0,\pm q_a,q_0), \quad \{ q_a \} \ne 0,
\\
\Omega(\gamD{1})=&\, 2\gfinv_{0}\quad\mbox{\rm for}\quad \gamD{1}=(0,0,0,q_0).
\end{split}
\ee
It is useful to note also that the contributions to \eqref{defG1} with $m=0$ are nothing
but the $\alpha'$-corrected part of the prepotential, $\sum_{n>0}\hH_{0,n}^{\rm D1}=F^{\alpha'\text{-loop}}+F^{\ws}$.
It is easy to check that both the new contours and contact Hamiltonians satisfy \eqref{mappatches} and \eqref{transhH}, respectively,
where in the last relation it is important to take into account the possibility to drop regular terms (see footnote \ref{foot-regular}).

\subsection{D5-instantons after gauge transformation}
\label{subsec-D5trans}

Performing the gauge transformation which puts D1-D(-1)-instantons into an S-duality invariant formulation, we rotated their
BPS rays to the real axis. On the way they will necessarily cross the other BPS rays of D3 and D5-branes.
Since the charges of the crossing rays are generically mutually non-commuting, i.e. $\langle\gamma,\tilde\gamma\rangle\ne 0$,
the gauge transformation should have a non-trivial effect on the transition functions of the other branes.

\lfig{Schematic representation of the twistor data generating D(-1)-D1 and
D5-instantons in the type IIB picture.}{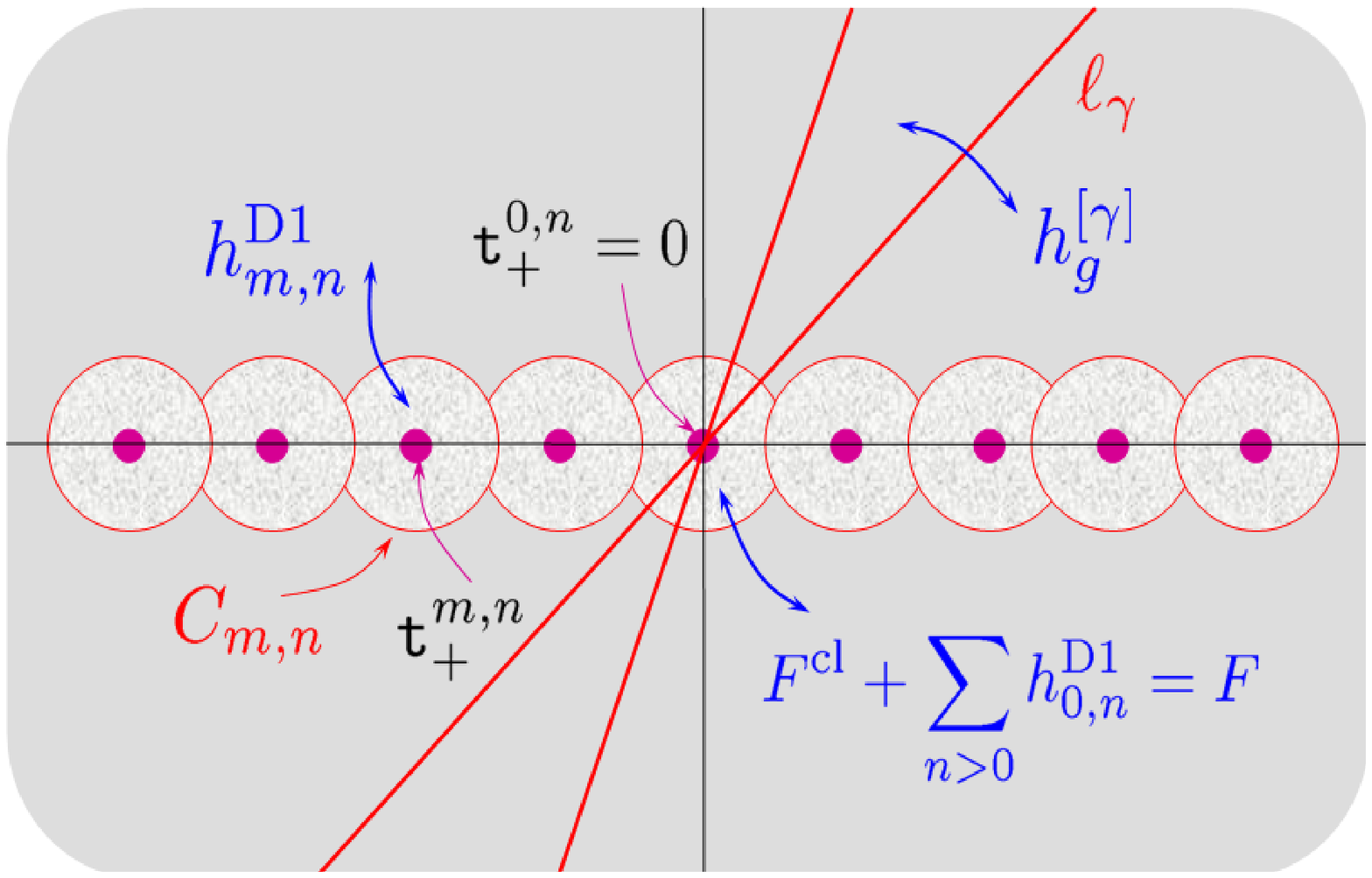}{12cm}{fig-D1inst}{-0.6cm}

Indeed, the general action of the gauge transformation is shown in \eqref{gaugetr}.
Assuming that we are at the point in the moduli space which does not belong to any line of marginal stability,
for $\gamma=(p^\Lambda, q_\Lambda)$ with non-vanishing $p^\Lambda$ the holomorphic functions \eqref{fungengauge}
generating the gauge transformation on the two sides of the BPS ray $\ell_\gamma$ will be the same.
Therefore, we turn out to be in the situation where the formula \eqref{gaugehHxi} can be applied.
As a result, the gauge transformed D5-brane contact Hamiltonians read
\be
\hHij{\gamma}_g(\xi,\txi)=H_{\gamma}(\Xi^{(g)}_\gamma),
\qquad
\Xi^{(g)}_\gamma=  \Xi_\gamma+p^\Lambda\p_{\xi^\Lambda}\gi{\gamma}(\xi).
\label{Hgam-tr}
\ee
The corresponding gauge transformed transition functions are more complicated and can be found in \eqref{trHTijall}.
All twistor data after the gauge transformation and the resummation are shown in Fig. \ref{fig-D1inst}.

The modification of the contact Hamiltonians \eqref{Hgam-tr} is crucial for keeping consistency of the twistor construction
with wall-crossing. Indeed, let us consider, for instance, the wall of marginal stability corresponding to the alignment of central charges
of a D5-brane of charge $\gamma$ and D(-1)-branes, so that after crossing this wall all BPS indices
$\bar\Omega(n\gamma+\gamD{-1})$, with $\gamD{-1}\in\GamD{-1}_\gamma$, change.
Since the central charges of D(-1)-branes are real, at the wall
the five-brane BPS rays $\ell_{n\gamma+\gamD{-1}}$ become aligned with the imaginary axis.
Before the gauge transformation, the D(-1) BPS rays $\ell_{\gamD{-1}}$ belonged to this axis and therefore, after crossing the wall,
the relative positions of $\ell_{n\gamma+\gamD{-1}}$ and $\ell_{\gamD{-1}}$ were exchanged.
This exchange compensated the change in the BPS indices
and ensured the smoothness of the contact structure and the metric on the moduli space across the wall.
But after the gauge transformation the contours associated with D(-1)-branes are rotated to the real axis.
Hence there is nothing to exchange its relative position with $\ell_{n\gamma+\gamD{-1}}$ to compensate the change of the BPS indices!
So how can the metric be still smooth in this gauge transformed picture?
It turns out that the smoothness is ensured precisely by the shift of $\Xi_\gamma$ in \eqref{Hgam-tr} induced by the gauge transformation.
The point is that the functions $\gi{\gamma}$ determining this shift are different on the different sides of the wall.
(In the considered example they differ by an overall sign due to the prefactor in \eqref{fungengauge}.)
As a result, crossing the wall, one also changes the form of the gauged transformed contact Hamiltonians, and
it is done in such a way that the combined effect of all changes is the smoothness of the moduli space.
A rigorous proof of this fact can be obtained by representing the gauge transformed KS operators as in \eqref{gaugetr}
and using the original KS wall-crossing formula \eqref{ewall}.

\section{Fivebrane instantons from S-duality}
\label{sec-fivebrane}

Now we have all ingredients to reach our main goal --- the twistorial description of fivebrane instantons in the presence of
D1-D(-1)-instanton corrections.\footnote{We remind that our construction ignores the effect of D3-instantons.
Although such approximation is physically unjustified, at a formal level it can be achieved by setting to zero all DT-invariants
$\Omega(\gamma)$ for charges with $p^0=0,\ p^a\ne 0$. Note however that we do include the effect of D3-branes bound to D5-branes,
as required by invariance under monodromies.}
To this end, we simply apply the modular constraint \eqref{transhH}
to the gauge transformed contact Hamiltonians \eqref{Hgam-tr} which are identified with the elements of an $SL(2,\IZ)$ multiplet with $m=0$.
More precisely, we set $\hHij{\hgam}_{0,p^0}=\hHij{\gamma}_g$ where we split charge $\gamma$ of a D5-D3-D1-D(-1)-bound state
into the D5-component $p^0$ and the reduced charge vector $\hgam=(p^a,q_a,q_0)$ identified with the index $i$ in \eqref{transhH}.
On this function we act by an $SL(2,\IZ)$ transformation parametrized as
\be
\label{Sdualde}
\gl{}= \begin{pmatrix} a & b \\ k/p^0 & p/p^0 \end{pmatrix} \in SL(2,\IZ)\, ,
\ee
where the two integers $(p,k)\neq (0,0)$ have $p^0$ as the greatest common divisor, whereas $a$ and $b$ must satisfy $a p - b k = p^0$.
The integer $k$ will appear as NS5-brane charge. As for the other charges, it is convenient to pack them into
rational charges $n^a=p^a/k$, $n^0=p/k$ and the so-called invariant charges \cite{Alexandrov:2010ca}
\be
\begin{split}
\hat q_a = &\, {q_a  + \frac12 \,\kappa_{abc} \frac{p^b p^c}{p^0},}
\\
\hat q_0 =&\, { q_0  + \frac{p^a q_a}{p^0} + \frac13\, \kappa_{abc}\frac{p^a p^b p^c}{(p^0)^2}},
\end{split}
\ee
which are invariant under the spectral flow transformation, whose action on the charge vector $\gamma$ is identical to
the action \eqref{bjacr} on the symplectic vector $(\zeta^\Lambda, \tzeta_\Lambda)$.

The $SL(2,\IZ)$ action on the contact Hamiltonian is easily computed using \eqref{SL2Zxi}.
Then the S-duality constraint implies that
\be
\begin{split}
\hkp=&\,(p^0)^{-1}(k\xi^0+p)\,\, \gl{}\cdot \hHij{\gamma}_g(\xi,\txi)
\\
=&\, \frac{\bar\Omega_{k,p}(\hgam)}{4\pi^2} \frac{k}{p^0} (\xi^0+n^0)\sigma_D(\gamma)\, \expe{ S_{k,p;\hgam}},
\end{split}
\label{fivebraneh}
\ee
where the result is written using the following notations:
\begin{itemize}
\item
fivebrane twistorial action
\be
\begin{split}
S_{k,p;\hgam}= &\, -k S_{n^\Lambda}+ \frac{p^0(p^0 \hat q_0-k \hat q_a (\xi^a + n^a))}{k^2(\xi^0 +n^0)}
- \frac{a}{k}\,p^0 q_0- c_{2,a} p^a \varepsilon(\gl{})
\\
&\,  -\frac{(-1)^{\Nq_{k,p}(\hgam)}}{2\pi\I}  \sum_{\gamD{1}\in\Gamma_{k,p;\hgam}}\bn_{\tilde q}\, p^\Lambda \tilde q_\Lambda \,
\expe{\tS_{k,p;\gamD{1}}}
\end{split}
\label{fivebraneaction}
\ee
with $S_{n^\Lambda} =\alpha - n^\Lambda \txi_\Lambda  + \Fcl (\xi + n)$;

\item
S-duality transformed D1-brane twistorial action\footnote{Note that both actions \eqref{fivebraneaction} and \eqref{Stildegam}
are regular at $k=0$ and reduce in this limit to the (gauge transformed) D-instanton twistorial actions $-\Xi^{(g)}_\gamma$ and $-\Xi_{\tilde\gamma}$,
respectively.}
\be
\tS_{k,p;\gamD{1}}= \frac{\tilde q_0 (p^0)^2}{k^2 (\xi^0 + n^0)}-\frac{p^0\tilde q_a \xi^a}{k(\xi^0+n^0)}-\frac{a}{k}\,p^0 \tilde q_0;
\label{Stildegam}
\ee

\item
rational Gopakumar-Vafa invariants $\bn_{q}$ constructed from $\gfinv_{q_a}$  as in \eqref{rationalinvariants};

\item
transformed BPS indices
$\bar\Omega_{k,p}(\hgam)= \bar\Omega(\gamma;\gl{}\cdot z)$ which take into account the fact that DT invariants
are only piecewise constant;

\item
transformed set of charges
\be
\Gamma_{k,p;\hgam}=
\GamD{1}_\gamma(\gl{}\cdot z)\cup \GamD{-1}_\gamma(\gl{}\cdot z),
\label{transGam}
\ee
where the dependence on $z^a$ comes from the dependence of \eqref{latDm1} and \eqref{latD1} on the central charge function,
or, more precisely, on the chamber in the moduli space, and is analogous to the dependence of the BPS indices;

\item
target quadrant in the complex plane
$\Nq_{k,p}(\hgam)= \left\lfloor\frac{2}{\pi}\, \arg \(\I \gl{}\cdot Z_\gamma\)\right\rfloor$.

\end{itemize}
The associated contours on $\CP$ are also just the images of $\ell_\gamma$ under \eqref{Sdualde} and thus can be written as
\be
\ell_{k,p;\hgam}=\{\varpi:\ Z_\gamma(\gl{}\cdot z)/(\gl{}\cdot t)\in\I\IR^-\}.
\label{BPSray-five}
\ee
From \eqref{SL2varpi} it follows that they are rays joining the points $\htpm^{k,p}$ (see Fig. \ref{fig-fiveinst}).
Together $\hkp$ and $\ell_{k,p;\hgam}$ determine the twistorial data sufficient to incorporate all fivebrane
instanton corrections to the metric on the HM moduli space.

\lfig{Schematic representation of the twistor data generating D(-1)-D1 and
all fivebrane instantons. BPS rays joining different points $\htpm^{k,p}$ correspond to different fivebrane charges $k$ and $p$.
Different BPS rays joining the same points correspond to different reduced charges $\hgam$.}{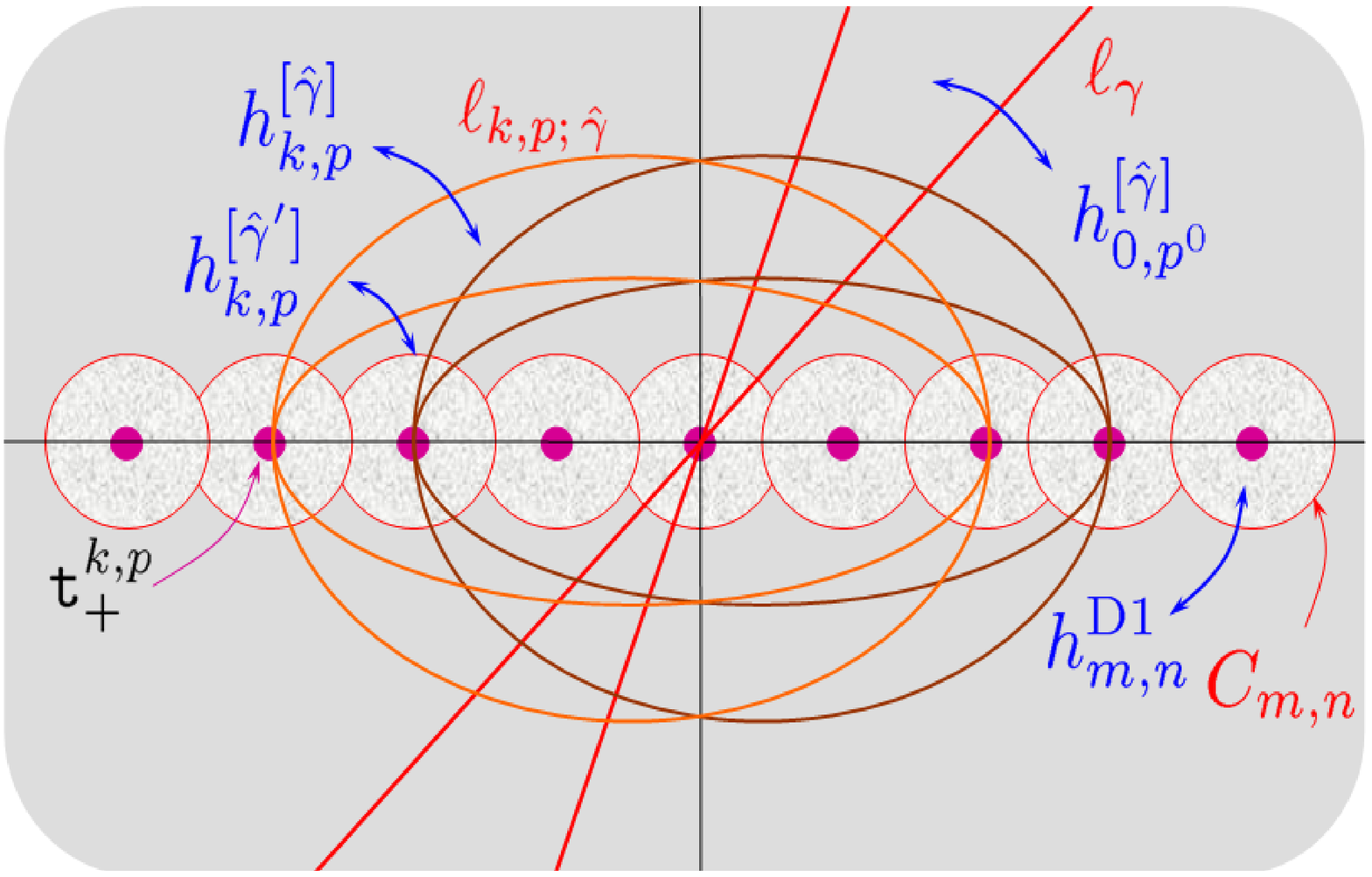}{12cm}{fig-fiveinst}{-0.6cm}

The function \eqref{fivebraneh} is almost identical to the result for the fivebrane transition function found in \cite[Eq.(5.30)]{Alexandrov:2010ca}
in the one-instanton approximation. It differs only by a prefactor ensuring the correct modular weight and by
the last term in \eqref{fivebraneaction} appearing as a result of the gauge
transformation \eqref{fungengauge}.\footnote{We also flipped the sign of the NS5-brane charge $k$.}
In particular, in \cite{Alexandrov:2010ca} it was shown that the saddle point evaluation of the Penrose transform
of this function yields the exponential of the NS5-brane instanton action found previously from
the analysis of classical supergravity solutions \cite{deVroome:2006xu}.

It is important however that, in contrast to \cite{Alexandrov:2010ca}, our result provides fivebrane
instanton corrections to the HM metric to {\it all orders} of the instanton expansion.
This expansion can be seen explicitly when one computes the contact transformation \eqref{newglue} generated by
\eqref{fivebraneh}. Equivalently, this calculation provides expressions for the corresponding transition function $\Hij{\hgam}_{k,p}$
and its derivatives. The former is given by
\be
\begin{split}
\Hij{\hgam}_{k,p}
=&\, \hkp+2\pi^2(\hkp)^2\(\frac{\hat q_0 (p^0)^2}{k(\xi^0+n^0)}+\frac{2k^2\Fcl(\xi+n)}{(1+2\pi\I k\hkp)^2} \)
\\
&\,
-(-1)^{\Nq_{k,p}(\hgam)}\,\frac{k(\xi^0+n^0)}{4\pi^2 p^0}\sum_{\gamD{1}\in\Gamma_{k,p;\hgam}}
\bn_{\tilde q}\,\expe{\tS_{k,p;\gamD{1}}} \cE\(\frac{4\pi^2 p^0 (\tilde{q}_\Lambda p^\Lambda)}{k(\xi^0+n^0)}\,\hkp\),
\end{split}
\label{tranNS5all}
\ee
where we introduced the function
\be
\cE(x)=1-(1+x)\, e^{-x},
\label{fun}
\ee
whereas the results for derivatives are reported in appendix \ref{ap-derfivebrane}.
They can be used to write down explicitly a system of integral equations which will provide a manifestly S-duality invariant
twistorial formulation of the HM moduli space including all D(-1), D1 and fivebrane instanton corrections.
Of course, this system cannot be solved analytically, but it should allow a perturbative solution generating the instanton expansion
around the classical metric.

A very non-trivial consistency check of our computation is that, as shown in appendix \ref{ap_Sdualconstrap},
the transition functions \eqref{tranNS5all} satisfy the non-linear S-duality constraint derived in \cite{Alexandrov:2013mha}.
It is amazing to see how all non-linearities fit each other, but it is even more remarkable that all of them
disappear once one starts working in terms of contact Hamiltonians.

Another consistency check is to verify that our results for fivebrane corrections are compatible with the action of all discrete isometries on $\cM_H$
which we presented in section \ref{subsec-Udual}. This is particularly important as in \cite{Alexandrov:2010ca}
it was found that there is a clash between the one-instanton approximation to fivebrane corrections, which is essentially identical
to our results, and the Heisenberg and monodromy symmetries. But as we argued, the monodromy transformations need to be modified
to ensure the correct group representation and it is natural to expect that this should resolve the above issue as well.
Indeed, due to the invariance of D-instanton corrections, the invariance of the contact structure affected by fivebrane instantons
is {\it guaranteed} by the closure of the group action.
Nevertheless, we demonstrate this invariance explicitly in appendix \ref{ap-MHinv}.

Finally, it is worth to note that S-duality generates a new family of walls in the moduli space $\cM_H$ which do not belong
to the \kahler moduli subspace $\SK$. These are the images of the original walls of marginal stability under S-duality transformation.
Since $z^a$ is mapped into $cc^a+dd^a+\I|c\tau+d|t^a$,
the position of the new walls depends on the RR-fields $c^a$ and the complexified string coupling $\tau$.
Crossing such a wall, one changes the values of the transformed BPS indices $\bar\Omega_{k,p}(\hgam)$ which gives rise
to a potential discontinuity in the contact structure and the moduli space metric. However,
they both do remain continuous because the new twistorial data is just an image of the data which was already shown to be smooth.
The mechanism ensuring the smoothness is the same as in the end of section \ref{subsec-D5trans}.
Alternatively, this can be seen as a result of the change of the set $\Gamma_{k,p;\hgam}$ \eqref{transGam} which determines
the effect of D1-D(-1)-branes on the fivebrane instantons. Its change together with a rearrangement of fivebrane BPS rays
guarantees the smoothness.

\section{Discussion}
\label{sec-concl}

The main result of this paper is the twistorial construction of the HM moduli space $\cM_H$ of CY string vacua
in the type IIB picture which includes effects from fivebrane and D1-D(-1)-instantons.
In particular, the constructed fivebrane instantons generically have non-vanishing NS5-brane charge.
All non-perturbative corrections are encoded in the two sets of holomorphic functions, \eqref{defG1} found in \cite{Alexandrov:2009qq}
and \eqref{fivebraneh} derived here. These functions generate a system of integral equations which determine
Darboux coordinates on the twistor space and thereby the metric on $\cM_H$.

The key element of this construction was the use of the contact bracket formalism which provides
a new parametrization of contact transformations. The contact bracket was shown to satisfy the crucial property \eqref{trans-contact},
analogous to a similar property of the Poisson bracket, which ensures that the contact Hamiltonians $\hHij{ij}$,
encoding the geometry of a QK manifold in this twistor approach, transform {\it linearly} under all isometries.
In particular, this implies their linear transformation under S-duality \eqref{transhH}, which was used to derive
the contact Hamiltonians corresponding to fivebrane instantons.

Another important step was to improve the action of discrete isometries on $\cM_H$ at quantum level.
Namely, we found that the closure of the duality group requires a modification of certain symmetry transformations.
This adjustment had a double effect: not only it provided a consistent implementation of all symmetries,
but it also resolved a tension between fivebrane instantons
and monodromy and Heisenberg symmetries observed in \cite{Alexandrov:2010ca}.

However, the proposed modification of the monodromy action on the RR-fields raises the following problem.
Before the modification, it was given in \eqref{bjacr} and this seemingly complicated transformation
in fact follows from the definition of the RR-scalars in terms of the B-field and
the RR-potential $A^{\rm even}\in H^{\rm even}(\CY,\IR)$
\be
\label{ABze}
A^{\rm even}\, e^{-B}
= \zeta^0 - \zeta^a \omega_a - \tzeta_a \omega^a-\tzeta_0 \omega_{\CYm}
\ee
just by applying the shift of the B-field and keeping the potential fixed. Therefore, it is natural to ask whether
the modified transformation \eqref{bjacr-mod} can be generated in the same way. This would imply that
either the l.h.s. of \eqref{ABze} should be modified and acquires additional non-homogeneous (in $A^{\rm even}$) terms,
or the RR-potential transforms itself.
Since the new terms in \eqref{bjacr-mod} have their origin in the quadratic refinement, one might expect that
in both cases the corrections appear from some subtleties in the definition of the one-loop determinant
around the D-instanton background similar to the issues discussed, for instance, in \cite{Freed:1999vc}.

Returning to the fivebrane instantons, we note that the construction presented in this paper calls for two natural
extensions. First, it clearly misses the D3-brane contributions. As was indicated in the Introduction, the actual problem
is to find how the corresponding subset of D2-instantons on the type IIA side can be rewritten
in an S-duality invariant way. Unfortunately, this was not understood even in the linear (one-instanton)
approximation. Hopefully, once this problem is resolved at one-instanton level,
the contact bracket formalism will provide a fully non-linear solution.

The second extension is, in contrast, to map the fivebrane instantons found here in the type IIB picture
into the mirror type IIA formulation. What is non-trivial is that the resulting NS5-brane instanton corrections
should be automatically symplectic invariant, a symmetry which is not seen on the type IIB side.
An interesting related question is whether these corrections will exhibit some form of integrability as
there are strong indications that the inclusion of NS5-instantons may be equivalent to quantization
of a certain integrable structure \cite{Alexandrov:2011ac,Alexandrov:2012np,Alexandrov:2013yva}.

The knowledge of fivebrane instantons also allows to approach two problems which are expected to be related to this type of
non-perturbative corrections. The first one is the existence of a singularity in the one-loop corrected metric on $\cM_H$.
This singularity should be resolved by non-perturbative effects, but D-instantons seem to be incapable to do so \cite{Alexandrov:2009qq}.
Thus, these are the NS5-brane corrections that should be responsible for the smoothness of the metric.
It will be a very non-trivial check on our construction to see whether the fivebrane instantons found in this paper
indeed resolve the singularity.

Another issue whose resolution was attributed to NS5-branes is the divergence of the sum over D-brane charges
appearing due to the exponential growth of the DT invariants \cite{Pioline:2009ia}.
Somehow NS5-brane effects should regulate this sum to make the non-perturbative metric on $\cM_H$ well defined.
It is likely however that solution to this problem requires the passage to the mirror type IIA picture,
which makes such a reformulation even more pressing.

Our final comment concerns the isometry group of $\cM_H$.
In this work it appears as a semidirect product of $SL(2,\IZ)$ with the nilpotent group
generated by the Heisenberg transformations and monodromies around the large volume point.
On the other hand, one might expect that the true U-duality group of the low energy theory
should be semisimple and is obtained by adding some new symmetry generators. Such extensions have been proposed
in \cite{Pioline:2009qt,Bao:2009fg,Persson:2011xi}, but it is not clear so far what can be such a group for generic CY.
It is interesting to see whether the contact bracket formalism can help solving this problem given that it is particularly
suited for dealing with symmetries.

\section*{Acknowledgments}

We are grateful to Davide Gaiotto, Albrecht Klemm, Jan Louis, Jan Manschot, Daniel Persson, Boris Pioline and Roberto Valandro
for valuable discussions and correspondence. We also thank Daniel Persson and Boris Pioline for careful reading of the manuscript.

\appendix

\section{Details on the isometry group}
\label{ap-Udual}

\subsection{The character $\varepsilon(\gl{})$ and the Dedekind sum}
\label{subap-Dedekind}

Before elucidating the group structure, we need to define the character $\varepsilon(\gl{})$ appearing
in the S-duality transformation of $\cla$ \eqref{cla}. It is given by the multiplier system of the Dedekind
eta function $\eta(\tau)$:
\be
\label{multeta}
e^{2\pi\I\, \varepsilon(\gl{}) }=\frac{\eta\left(\frac{a\tau+b}{c\tau+d}\right)}{(c\tau+d)^{1/2}\,
\eta(\tau)}.
\ee
In particular, $24\varepsilon(g)$ is an integer and $\varepsilon(\gl{})$ has the following explicit representation
\be
 \varepsilon(\text{g}) =
 \begin{cases}
      \frac{b}{24}\, \sign(d) & (c=0) \\
      \frac{a+d}{24c}-\hf s(d,c) -\frac18 & (c>0) \\
      \frac{a+d}{24c} + \hf s(d,c) + \frac18 & (c<0)
   \end{cases}
\ee
where $s(d,c)$ is the Dedekind sum. It can be written in terms of the sawtooth function
\be
\(\(x\)\)=\left\{\begin{array}{ll}
x-\lfloor x\rfloor -1/2, &\quad {\rm if \ }x\in \IR\setminus\IZ
\\
0, & \quad {\rm if \ } x\in \IZ
\end{array}\right.
\ee
as
\be
s(d,c) =  \sum_{r \text{ mod } |c|} \(\(\frac{r}{|c|}\)\) \(\(\frac{r d}{|c|}\)\).
\ee
An easy calculation leads to a more explicit expression
\be
s(d,c)=  \sum^{c-1}_{r=1}  \frac{r}{c} \(\(\frac{rd}{c}\)\)
= (c-1)\(\frac{d}{6c}\,(2c-1)-\frac14 \)-\sum^{c-1}_{r=1}  \frac{r}{c} \left\lfloor\frac{rd}{c}\right\rfloor,
\label{Dedekindsum}
\ee
where we set $c>0$.
Thus, for the generators \eqref{genST}, one obtains
\be
\varepsilon(S)=-\frac{1}{8},
\qquad
\varepsilon(T)=\frac{1}{24}.
\ee
In appendix \ref{ap-MHinv} we will also need the reciprocity relation satisfied by the Dedekind sum.
For coprime positive integers $d$ and $c$, it reads
\be
s(d,c)+s(c,d)=\frac{1}{12}\(\frac{d}{c}+\frac{1}{cd}+\frac{c}{d}\)-\frac14.
\label{reclaw}
\ee

\subsection{Failure of the group law}

We start by analyzing the group of discrete isometries presented in section \ref{subsubsec-isom}.
For our purposes it is convenient to express the action of their generators in the type IIB coordinate basis.
The result is given in Table \ref{tab-Unon} where we use the notations introduced in section \ref{subsec-correcttr}.

\begin{table}
\vspace{0.cm}\hspace{-1.7cm}
\begin{tabular}{|c|c|c|c|c|c|}
\hline $\vphantom{\frac{A^{A^A}}{A_{A_A}}}$
& $b^a$ & $ c^a$ & $\cla$ &$\cl0$ & $\psi$
\\
\hline $\vphantom{\frac{A^{A^A}}{A_{A_A}}}$
$S$ &  $c^a$     &  $-b^a$  & $\cla+\frac{c_{2,a}}{8}$
& $-\psi$ & $\cl0$
\\
\hline $\vphantom{\frac{A^{A^A}}{A_{A_A}}}$
$T$ &  $b^a$     &  $c^a+b^a$  & $\cla-\frac{c_{2,a}}{24}$
& $\cl0$ & $\psi-\cl0$
\\
\hline $\vphantom{\frac{A^{A^A}}{A_{A_A}}}$
$\opT^{(0)}_{\eta^0}$ &   $b^a$    &   $c^a+\eta^0 b^a$  & $\cla-\frac{c_{2,a}}{24}\,\eta^0$
& $\cl0$
& $\psi-\eta^0\cl0+\eta^0\phi_0$
\\
\hline $\vphantom{\frac{A^{A^A}}{A_{A_A}}}$
$\opT^{(1)}_{\epsilon^a,0}$ &   $b^a+\epsilon^a$    &   $c^a$  & $\cla+\frac12\, \kappa_{abc} \epsilon^b c^c$
& $\begin{array}{c}\cl0-\epsilon^a \cla \\-\frac16\, \kappa_{abc} \epsilon^a (b^b+2\epsilon^b)c^c\end{array}$
& $\psi+\frac16\, \kappa_{abc}\epsilon^a c^b c^c- \kappa_a \epsilon^a$
\\
\hline $\vphantom{\frac{A^{A^A}}{A_{A_A}}}$
$\opT^{(1)}_{0,\eta^a}$&  $b^a$    &    $c^a+\eta^a$  & $\cla-\frac12\, \kappa_{abc} \eta^b b^c+A_{ab} \eta^b$
&  $\cl0+\frac16\, \kappa_{abc} \eta^a b^b b^c +\frac{c_{2,a}}{24}\, \eta^a $
&  $\begin{array}{c}\psi+\eta^a (\cla-\phi_a)+\frac12\, A_{ab} \eta^a \eta^b\\
-\frac16\, \kappa_{abc}\eta^a b^b (c^c+2\eta^c)\end{array}$
\\
\hline $\vphantom{\frac{A^{A^A}}{A_{A_A}}}$
$\opT^{(2)}_{\tilde\eta_a}$
&  $b^a$  & $c^a$ & $\cla+\tilde\eta_a$
& $\cl0$  & $ \psi$
\\
\hline $\vphantom{\frac{A^{A^A}}{A_{A_A}}}$
$\opT^{(3)}_{\tilde\eta_0,\kappa}$ &  $b^a$  & $c^a$ & $\cla$
&  $\cl0+\tilde\eta_0$ & $\psi+\kappa$
\\
\hline
\end{tabular}
\caption{The action of generators of the discrete symmetry transformations in the type IIB coordinate basis before modifications.}
\label{tab-Unon}
\end{table}
\vspace{0.cm}

Already a quick glance on the table reveals the first problem. Comparing the action of the $T$-generator of $SL(2,\IZ)$ and
the Heisenberg shift $\opT^{(0)}_{1}$, one finds that they are almost identical in agreement
with the expectation that they represent the same symmetry transformation.
But this identification works only if one sets the characteristic $\phi_0$ to zero.\footnote{The difference
cannot be compensated by the other Heisenberg shift $\opT^{(3)}_{(0,\kappa)}$ since $\kappa\in\IZ$ whereas $\phi_0$ is generically non-integer.}
However, this restriction seems to be inconsistent with the transformation property \eqref{transchar} of $\phi_0$ under monodromies.

An even more serious problem appears when one considers the commutator of the Heisenberg transformation $\opT^{(1)}_{(0,\eta^a)}$
with the generator $S$. A straightforward calculation shows that
\bea
S^{-1}\, \opT^{(1)}_{0,\eta^a}\, S\cdot
\(\begin{array}{c}
b^a \\ c^a \\ \cla \\ \cl0 \\ \psi \end{array}\)
&=&
\opT^{(1)}_{\eta^a,0}\cdot
\(\begin{array}{c} b^a \\ c^a \\ \cla \\ \cl0 \\ \psi \end{array}\)
+\(\begin{array}{c} 0 \\ 0 \\ A_{ab}\eta^b\\ \eta^a\(\phi_a + \frac{c_{2,a}}{8}-\hf\, A_{ab}\eta^b\)
\\ \eta^a\(\kappa_a+\frac{c_{2,a}}{24}\) \end{array}\).
\label{twistheisen}
\eea
Thus, the commutator acquires an anomalous contribution which cannot be produced by the action of other generators and therefore
is inconsistent with the group structure. Although the anoomalous contributions to the transformation of $\cl0$ and $\psi$
can be, in principle, canceled by the appropriate choice of the characteristic $\phi_a$ and the character $\kappa_a$,
the anoamly in the transformation of $\cla$ is not removable.
A similar anomaly arises also in the commutator of $S$ with the monodromy transformation.
Hence, we have to conclude that the transformations displayed in Table \ref{tab-Unon}
fail to form a representation of the duality group of the theory.

\subsection{Improved duality group}

Here we list the group relations for the symmetry generators from Table \ref{tab-U} obtained by the modification
explained in section \ref{subsec-correcttr}. They explicitly demonstrate that the improved transformations provide
a nice group representation. To display the results, we use the notation $[A,B]=A^{-1}B^{-1} A B$.
\begin{itemize}
\item
{Commutators with $S$:}
\be
\begin{split}
S^{-1} \,\opT^{(1)}_{\epsilon^a,0}\, S=&\,\,
\opT^{(1)}_{0,-\epsilon^a}\,\opT^{(2)}_{2 A_{ab} \epsilon^b}\, \opT^{(3)}_{0,\,3L_0(\epsilon) - \epsilon^a L_a(-\epsilon)},
\\
S^{-1}\, \opT^{(1)}_{0,\eta^a}\, S=&\,\,
\opT^{(1)}_{\eta^a,0},
\\
S^{-1}\, \opT^{(2)}_{\tleta_a} \,S =&\,\,
\opT^{(2)}_{\tleta_a},
\\
S^{-1}\, \opT^{(3)}_{\tleta_0,\kappa}\, S=&\,\,
\opT^{(3)}_{-\kappa,\tleta_0};
\end{split}
\label{monS}
\ee

\item
{Commutators with $T$:}
\be
\begin{split}
[\opT^{(1)}_{\epsilon^a,0},T] =&\,\,
\opT^{(1)}_{0,\epsilon^a}\opT^{(2)}_{L_a(\epsilon)}
\opT^{(3)}_{-L_0(\epsilon^a),\,  \epsilon^a L_a(\epsilon) - L_0(\epsilon)},
\\
[ \opT^{(1)}_{0,\eta^a},T]=&\,\,
[\opT^{(2)}_{\tilde\eta^a},T]=\unit,
\\
[\opT^{(3)}_{\tleta_0,\kappa},T] =&\,\,
\opT^{(3)}_{0,-\tleta_0 };
\end{split}
\label{commT}
\ee

\item
{The nilpotent subgroup:}
\be
\begin{split}
\[\opT^{(1)}_{\epsilon^a,0}, \opT^{(1)}_{0,\eta^a} \]
= &\,\,
\opT^{(2)}_{-\kappa_{abc}\epsilon^b\eta^c} \, \opT^{(3)}_{\eta^a L_a(-\epsilon),-\epsilon^a L_a(\eta)},
\\
\[\opT^{(1)}_{\epsilon^a,0}, \opT^{(2)}_{\tilde\eta_a} \]
=&\,\,
\opT^{(3)}_{\epsilon^a \tilde\eta_a,0}.
\\
\[\opT^{(1)}_{0,\eta^a}, \opT^{(2)}_{\tilde\eta_a} \]
=&\,\,
\opT^{(3)}_{0,-\eta^a\tilde\eta_a },
\\
\[\opT^{(1)}_{0,\eta^a}, \opT^{(3)}_{\tilde\eta_0,\kappa} \]=&\,
\[\opT^{(1)}_{\epsilon^a,0},\opT^{(3)}_{\tilde\eta_0,\kappa} \]=
\[\opT^{(2)}_{\tilde\eta_a},\opT^{(3)}_{\tilde\eta_0,\kappa} \]=\unit.
\end{split}
\ee
\end{itemize}

\section{Transformation property of the contact bracket}
\label{ap-contactbr}

The aim of this appendix is to present a proof of the transformation property \eqref{trans-contact}
of the contact bracket under contactomorphisms rescaling the contact one-form $\cX$ by a holomorphic factor $\lambda$.
It seems to us more instructive to give this proof in a coordinate independent way. In this language the contact bracket is defined
as $\{h, f\}=\Xf{h}(f)$, whereas the vector field $\Xf{h}$ is determined by the following properties
\be
\ii_{\Xf{h}}\de \cX=-\de h+ R(h) \cX,
\qquad
\ii_{\Xf{h}}\cX=h,
\label{defXh}
\ee
where $\ii_X$ denotes the contraction of the vector $X$ with a differential form and
$R=\p_\alpha$ is the Reeb vector field which is the unique element of the kernel of $\de\cX$ such that $\cX(R) =1$.
To prove \eqref{trans-contact}, it is sufficient to show that
\be
\vrh \cdot \Xf{h}=\Xf{\lambda^{-1} \vrh\cdot h}.
\ee

To simplify notations, let us denote by prime the transformed quantities.
Then we need to prove that
\be
\ii_{\Xf{\lambda^{-1}h'}}\de\(\lambda \cX\)=-\de h'+ R'(h') \lambda\cX,
\qquad
\ii_{\Xf{\lambda^{-1}h'}}\(\lambda\cX\)=h'
\label{defXh-prime}
\ee
provided $\Xf{h}$ satisfies \eqref{defXh} for any function $h$.
The second equality clearly holds since it boils down to the second equality in \eqref{defXh}.
On the other hand, the first one can be rewritten as
\be
\begin{split}
&\,
\(\ii_{\Xf{\lambda^{-1}h'}}\de\lambda\) \cX-\(\ii_{\Xf{\lambda^{-1}h'}}\cX\) \de\lambda+\lambda\,\ii_{\Xf{\lambda^{-1}h'}}\de\cX
+\de h'- \lambda  R'(h') \cX
\\
&\, =
\lambda\(\ii_{\Xf{\lambda^{-1}h'}}\de\log\lambda
+ R(\lambda^{-1}h')- R'(h')\) \cX=0.
\end{split}
\label{prop-to-prove}
\ee
Now let us apply the transformation $\vrh$ to the defining properties of the Reeb vector. This gives
\be
\cX(R')=\lambda^{-1},
\qquad
\ii_{R'}\de\(\lambda\cX\)=\(\ii_{R'}\de\lambda\)\cX-\de\log\lambda+\lambda\,\ii_{R'}\de\cX=0.
\ee
Contracting the second identity with $\Xf{\lambda^{-1}h'}$, one finds
\be
\begin{split}
&\,
\lambda^{-1}h' R'(\lambda)-\ii_{\Xf{\lambda^{-1}h'}}\de\log\lambda-\lambda\,\ii_{R'}\ii_{\Xf{\lambda^{-1}h'}}\de\cX
\\
&\, =R'(h')-\ii_{\Xf{\lambda^{-1}h'}}\de\log\lambda-R(\lambda^{-1} h')=0.
\end{split}
\ee
This shows that \eqref{prop-to-prove} indeed vanishes and completes the proof.

\section{Verifying S-duality constraint}
\label{ap_Sdualconstrap}

In \cite{Alexandrov:2013mha} it was shown that the transition functions describing a QK manifold carrying an isometric action of $SL(2,\IZ)$
should satisfy a non-linear constraint restricting their behavior under the $SL(2,\IZ)$ transformations.
Using notations from the end of section \ref{subsec-Sdual}, the constraint can be written as
\be
\begin{split}
\Hij{i}_{m,n}\ \mapsto\  &\, \frac{\Hij{i}_{m',n'}}{c\xii{\indlp}^0+d}
+ \frac{c}{6}\, \kappa_{abc} \frac{2 \Tmn{a}  - 3 \xii{\indlp}^a}{(c\xii{\indlp}^0 +d)(c\xii{\indrp}^0 +d)}\, \Tmn{b}\Tmn{c}
\\
&
+ \frac{c^2}{6}\, \kappa_{abc} \frac {\Tmn{a}\Tmn{b} +
3 (\xii{\indlp}^a - \Tmn{a})\xii{\indlp}^b }{(c\xii{\indlp}^0 +d)(c\xii{\indrp}^0 +d)^2}\, \Tmn{c}\Tmn{0}
\\
&
- \frac{c^3}{6}\, \kappa_{abc} \frac{\xii{\indlp}^a \xii{\indlp}^b\xii{\indlp}^c }{(c\xii{\indlp}^0 +d)^2 (c\xii{\indrp}^0 +d)^2}\, (\Tmn{0})^2,
\end{split}
\label{Sdualconstr}
\ee
where $\Tmn{\Lambda} \equiv \p_{\txii{\indrp}_\Lambda} \Hij{i}_{m',n'} - \xii{\indrp}^\Lambda \p_{\ai{\indrp}} \Hij{i}_{m',n'} $
and, to avoid cluttering, we denoted by ``$\indlp$" and ``$\indrp$" the two patches lying, respectively, on the left
and on the right of the contour $C_{m',n';i}$, omitting all the indices which these objects should in principle carry.
In this appendix we want to verify whether our results for fivebrane instanton corrections to the HM moduli space metric
are consistent with the constraint \eqref{Sdualconstr}. In principle, this should be guaranteed by the consistency
of the whole construction.
Thus, this appendix may be viewed as a non-trivial cross-check on the results reported in the main text.

First, what we need is the explicit expressions for the gauge transformed D-instanton transition functions $\Hij{\gamma}_g$
which will be identified with the $(0,n)$ elements of the $SL(2,\IZ)$ multiplet of fivebrane transition functions,
$\Hij{\hgam}_{0,p^0}=\Hij{\gamma}_g$.
The simplest way to get them is to apply the relation \eqref{relHH} where the corresponding contact Hamiltonian is given in \eqref{Hgam-tr}.
In this way, one finds
\bea
\begin{split}
\Hij{\gamma}_g =&\, \sum_{m=1}^\infty \frac{1}{m!}\[
\Xf{\hHij{\gamma}_g}^{m-1} \cdot \(1-\xi^\Lambda\p_{\xi^\Lambda}\)\hHij{\gamma}_g
+\xi^\Lambda\Xf{\hHij{\gamma}_g}^{m-1} \cdot\p_{\xi^\Lambda}\hHij{\gamma}_g \]
\\
=&\, \hHij{\gamma}_g +\sum_{m=2}^\infty \frac{\p_{\txi_\Lambda}\hHij{\gamma}_g }{m(m-2)!}\,
\Xf{\hHij{\gamma}_g}^{m-2} \cdot\p_{\xi^\Lambda}\hHij{\gamma}_g,
\end{split}
\label{evalHT}
\eea
where we used the properties of the contact bracket and the fact that $\Xf{\hHij{\gamma}_g}^{m} \cdot \xi^\Lambda=0$ for $m\ge 2$.
Furthermore, it is easy to see that for $m>2$ one has
\be
\begin{split}
\Xf{\hHij{\gamma}_g}^{m-2} \cdot\p_{\xi^\Lambda}\hHij{\gamma}_g=&\, \(-2\pi\I\hHij{\gamma}_g\)^{m-1}
\[\(p^\Sigma\p_{\xi^\Sigma}\)^{m-1} \p_{\xi^\Lambda}\gi{\gamma}\]
\\
=&\,
\frac{(-1)^{\Nq(\gamma)}}{2\pi\I}\sum_{\gamD{1}\in\GamD{-1}_\gamma\cup \GamD{1}_\gamma}
\tilde q_\Lambda\(-4\pi^2 \tilde q_\Lambda p^\Lambda \hHij{\gamma}_g\)^{m-1}\bn_{\tilde q}\,\expe{-\tilde q_\Lambda \xi^\Lambda}.
\end{split}
\ee
Substituting this into \eqref{evalHT} and making resummation over $m$, one arrives at the following result
\be
\Hij{\gamma}_g
= \hHij{\gamma}_g +2\pi^2 q_\Lambda p^\Lambda (\hHij{\gamma}_g)^2
-\frac{(-1)^{\Nq(\gamma)}}{4\pi^2}\sum_{\gamD{1}\in\GamD{-1}_\gamma\cup \GamD{1}_\gamma}
\bn_{\tilde q}\,\expe{-\tilde q_\Lambda \xi^\Lambda}\cE\(4\pi^2 \tilde q_\Lambda p^\Lambda \hHij{\gamma}_g\),
\label{trHTijall}
\ee
where $\cE(x)$ is defined in \eqref{fun}.

The next step is to evaluate the r.h.s. of \eqref{Sdualconstr}.
The functions $\Hij{\hgam}_{m,n}$ for $(m,n)=(k,p)$ are given in \eqref{tranNS5all}.
Their derivatives can be found in appendix \ref{ap-derfivebrane}.
In particular, the result for $\Tmn{\Lambda}$ \eqref{Tlambda} implies that
\be
\xii{\indrp}^\Lambda+n^\Lambda=\frac{\xii{\indlp}^\Lambda+n^\Lambda}{1+2\pi\I k\hkp}.
\label{relxiij}
\ee
Using these results together with the identity
\be
q_\Lambda p^\Lambda-\frac{2k^3}{p^0}\, n^0 \Fcl(n)=\hat q_0 p^0
\ee
and dropping again the patch indices of the Darboux coordinates, the r.h.s. of \eqref{Sdualconstr}
for $(m',n')=(k,p)$ and $(c,d)=(k/p^0,p/p^0)$ becomes
\be
\begin{split}
&\, \frac{p^0 \,\hkp}{k(\xi^0+n)}
+2\pi^2 q_\Lambda p^\Lambda\,\frac{ (p^0)^2(\hkp)^2 }{k^2(\xi^0+n^0)^2}
\\
&\,
-\frac{(-1)^{\Nq_{k,p}(\hgam)}}{4\pi^2}\sum_{\gamD{1}\in\Gamma_{k,p;\hgam}}
\bn_{\tilde q}\,\expe{\tS_{k,p;\gamD{1}}} \cE\(\frac{4\pi^2 p^0 (\tilde{q}_\Lambda p^\Lambda)}{k(\xi^0+n^0)}\,\hkp\).
\end{split}
\label{rhs-constr}
\ee
It is immediate to see that this expression is reproduced by applying the $SL(2,\IZ)$ transformation \eqref{Sdualde}
to the function \eqref{trHTijall}.
This completes the proof that the S-duality constraint \eqref{Sdualconstr} is indeed satisfied.

\section{Monodromy and Heisenberg invariance of fivebrane instantons}
\label{ap-MHinv}

In this appendix we check that the fivebrane corrections derived in section \ref{sec-fivebrane} preserve the isometric action

of the full duality group. The action of its generators on the coordinates on $\cM_H$ is presented in Table \ref{tab-U}.
It is clear that the contact structure on the twistor space and hence the metric on $\cM_H$
are invariant under the generators of the $SL(2,\IZ)$ subgroup by construction, and one should
check only its invariance under the nilpotent subgroup generated by $\opT^{(n)}$.

The invariance under the transformations with $n\ge 2$ is actually trivial. Indeed, their lift to the twistor space
acts on Darboux coordinates by simple shifts. The only non-trivial transformations are
\be
\begin{split}
\opT^{(2)}_{\tilde\eta_a}\ : \quad &\, \txi_a\ \mapsto\ \txi_a+\tleta_a,
\\
\opT^{(3)}_{\tilde\eta_0,-\kappa}\ : \quad &\,\txi_0\ \mapsto\ \txi_0+\tleta_0,
\quad
\alpha\ \mapsto\ \alpha+\kappa.
\end{split}
\ee
It is clear that they leave invariant both the contact Hamiltonians \eqref{fivebraneh} and the transition functions \eqref{tranNS5all}.

The check of the invariance under the first order generators is more non-trivial.
Let us start from the Heisenberg shift $\opT^{(1)}_{0, \eta^a}$ whose lift to the twistor space is given by
\be
\opT^{(1)}_{0, \eta^a}\ : \quad  \begin{array}{c}
\displaystyle{\xi^a\ \mapsto\ \xi^a-\eta^a,
\qquad
\txi_a\ \mapsto\ \txi_a+A_{ab}\eta^b,
\qquad
\txi_0\ \mapsto\ \txi_0+\frac{c_{2,a}}{24}\,\eta^a,}
\\
\displaystyle{\alpha\ \mapsto\ \alpha+\eta^a\txi_a+\hf\, A_{ab}\eta^a\eta^b.}
\end{array}
\label{Heisact-Darboux}
\ee
It is easy to see that the contact Hamiltonians generating fivebrane instanton corrections are {\it not} invariant under this action.
However, the invariance can be restored if one makes a compensating transformation of the charges $\hgam\mapsto \hgam[\eps]$:
\be
\label{spectrflow}
\begin{array}{c}
\displaystyle{
p^a[\eps]= p^a + \epsilon^a p^0,
\qquad
q_a[\eps]= q_a -\kappa_{abc}p^b \epsilon^c
-\frac{ p^0}2\,\kappa_{abc} \epsilon^b \epsilon^c ,}
\\
\displaystyle{q_0[\eps]= q_0 -q_a \epsilon^a+\frac12\, \kappa_{abc}p^a \epsilon^b \epsilon^c
+\frac{ p^0}6\,\kappa_{abc} \epsilon^a \epsilon^b \epsilon^c}
\end{array}
\ee
with parameter $\epsilon^a=k\eta^a/p^0$,
which is nothing else but the spectral flow transformation (cf. \eqref{bjacr}).
In other words we claim that
\be
\opT^{(1)}_{0, \eta^a}\cdot \hHij{\hgam}_{k,p}= \hHij{ \hgam[-k\eta^a/p^0]}_{k,p},
\qquad
\opT^{(1)}_{0, \eta^a}\cdot \ell_{k,p;\hgam}= \ell_{k,p;\hgam[-k\eta^a/p^0]},
\label{Heisinv}
\ee
which ensures that the contact structure stays invariant and therefore the transformation is an isometry of the moduli space.

The second equality in \eqref{Heisinv} follows from two facts. First, since the transformation \eqref{Heisact-Darboux}
does not affect $\xi^0$ and the fiber coordinate $t$, the points $\htpm^{m,n}$ and hence $\gl{}\cdot t$ stay invariant.
Second, as one can check, the combined action of $\opT^{(1)}_{0, \eta^a}$ and the spectral flow
\eqref{spectrflow} leaves invariant the S-duality transformed central charge $Z_\gamma(\gl{}\cdot z)$.
Then the above transformation of the contours $\ell_{k,p;\hgam}$ immediately follows from their definition \eqref{BPSray-five}.

To prove the first equality in \eqref{Heisinv}, we need a few more observations:
\begin{itemize}
\item
The transformation of the transformed BPS indices $\bar\Omega_{k,p}(\hgam)$ under the combined action reads
\be
\opT^{(1)}_{0, \eta^a}\cdot \bar\Omega_{k,p}(\hgam[\eps])= \bar\Omega(\gamma[\eps];\gl{}\cdot z+\eps)
=\gl{}\cdot\bar\Omega(\gamma[\eps]; M_{\eps^a}\cdot z).
\ee
But the rational DT invariants are known to be invariant under the combination of monodromy and
spectral flow transformations with the same parameters \cite{Manschot:2009ia}.
As a result, one obtains the invariance of $\bar\Omega_{k,p}(\hgam)$.

\item
Performing the change of the summation variable in the last term of $S_{k,p;\gamma}$ \eqref{fivebraneaction},
\be
\tilde q_0\ \mapsto\ \tilde q_0 - \epsilon^a \tilde q_a,
\qquad
\tilde q_a\  \mapsto\ \tilde q_a
\label{comptildeq}
\ee
with the same $\epsilon^a$ as above, one finds that this term is
invariant as well. Indeed, both the prefactor $p^\Lambda\tilde q_\Lambda$ and the exponential of $\tS_{k,p;\gamD{1}}$ stay invariant.
One could worry that the change of $\tilde q_0$ affects the set of charges one sums over, but in fact
it just compensates the change in $\Gamma_{k,p;\hgam}$ induced by the transformation of $\hgam$ and the moduli:
this set of charges  is defined in terms of
the S-duality transformed central charges $\gl{c,d}\cdot Z_\gamma$ and $\gl{c,d}\cdot Z_{\gamD{1}}$ which are both
invariant under the combination of the three transformations $\opT^{(1)}_{0, \eta^a}$, \eqref{spectrflow} and \eqref{comptildeq}.
\end{itemize}
Finally, one can see that the invariance of the combination  $\xi^\Lambda+n^\Lambda$ and the charges $\hat q_\Lambda$
ensures that the whole function $\hHij{\hgam}_{k,p}$ \eqref{fivebraneh} transforms at most by a phase independent of the fields,
$
\opT^{(1)}_{0, \eta^a}\cdot \hHij{ \hgam[\eps]}_{k,p}= \nu(\eta) \hHij{ \hgam}_{k,p}
$.
A direct calculation leads to the following expression
\bea
\begin{split}
 \nu (\eta)=&\,
\expe{\frac{k}{2p^0}\, \kappa_{abc} p^a p^b \eta^c +\frac{k}{2}\,c_{2,a}\eta^a \[s\(\frac{p}{p^0},\frac{k}{p^0}\)
-\frac{1}{4}\(p^0-1\)\]
\right.
\\
& \left.\qquad
-\frac{k(k-1)}{2}A_{ab}\eta^a\eta^b
+\eta^a A_{ab} p^b -\frac{ap^0}{k}\,p^\Lambda L_\Lambda(\epsilon)}.
\end{split}
\eea
Using the properties \eqref{propA} and \eqref{c2aprop}, this can be rewritten as
\be
\nu(\eta)
= \expe{  p^0 c_{2,a} \eta^a  \[ \frac{c}{2}\, s\(d,c \)
- \frac{c p^0}{8}\(1-c\) +\frac{a }{12} \(c^2 -1\) \]
-(a-1) \(c+ 1 \) A_{ab} \eta^a p^b },
\label{Hphase}
\ee
where we preferred to write the result in terms of $c=k/p^0$ and $d=p/p^0$.
Now note that the relation $ad - bc =1 $ ensures that $a$ and $c$ can not be simultaneously even.
Therefore, the last term in \eqref{Hphase} is an integer and thus disappears.
Furthermore, using the expression for the Dedekind sum \eqref{Dedekindsum} and taking into account that $c_{2,a} \eta^a$ is even,
one finds
\be
\nu(\eta) = \expe{ p^0 c_{2,a} \eta^a  \(c -1 \)\[
\frac{c }{8}\,(p^0-1)+\frac{d}{12}\(2c -1 \)+\frac{a }{12}\,(c+ 1 )\] }.
\label{Heisen1}
\ee
Since $\frac{1}{4}\,c(c -1 ) p^0(p^0-1)\in\IZ$, the first term can actually be dropped.
On the other hand, inserting in the last term
$1=ad-bc$ and using that $\frac16\,c(c-1)(c+1)\in \IZ$ as well as $\hf\, c(c-1)\in \IZ$, one arrives at
\be
\nu(\eta) = \expe{ \frac{d}{12}\,c_{2,a} \eta^a (c+1)(c-1)(a-1)(a+1)}.
\ee
But it is easy to realize that $(c+1)(c-1)(a-1)(a+1)$ is divisible by 6. Indeed, since $a$ and $c$ cannot be simultaneously even,
the above expression is divisible by 2. Similarly, it is not divisible by 3 only if $c=3m$, $a=3n$, which is in
contradiction with $ad-bc=1$. Thus, the expression in the exponential is an integer and the phase is trivial $\nu(\eta)=1$.
This completes the proof of \eqref{Heisinv}.

The last generator to be checked is $\opT^{(1)}_{\eps^a,0}$ corresponding to monodromy transformations.
Its action lifts to the twistor space as follows
\be
\opT^{(1)}_{\eps^a,0}\ :\quad
\begin{array}{rl}
\xi^0\ \mapsto\ & \displaystyle{\xi^0,
\qquad
\xi^a \ \mapsto\ \xi^a + \epsilon^a \xi^0,}
\\
\txi_a\ \mapsto\ & \displaystyle{\txi_a - \kappa_{abc} \epsilon^b \xi^c - \frac12\, \kappa_{abc} \epsilon^b \epsilon^c \xi^0
+A_{ab}\epsilon^b,}
\\
\txi_0\ \mapsto\ & \displaystyle{\txi_0 - \epsilon^a \txi_a + \frac12\, \kappa_{abc} \epsilon^a\epsilon^b \xi^c
+ \frac16\, \kappa_{abc} \epsilon^a\epsilon^b\epsilon^c \xi^0
-\hf\, A_{ab}\epsilon^a\epsilon^b+\frac{c_{2,a}}{8}\, \epsilon^a,}
\\
\alpha \ \mapsto\ &\displaystyle{ \alpha + \frac12 \( \kappa_{abc} \epsilon^a \xi^b\xi^c +
\kappa_{abc} \epsilon^a\epsilon^b\xi^c \xi^0
+ \frac13\, \kappa_{abc} \epsilon^a \epsilon^b\epsilon^c (\xi^0)^2 \) +\frac{c_{2,a}}{24}\, \eps^a.}
\end{array}
\label{monDc}
\ee
Similarly to the relation \eqref{Heisinv}, the twistor data generating fivebrane instantons can be shown to satisfy
\be
\opT^{(1)}_{\eps^a,0}\cdot \hHij{\hgam}_{k,p}= \hHij{ \hgam[-p\eps^a/p^0]}_{k,p},
\qquad
\opT^{(1)}_{\eps^a,0}\cdot \ell_{k,p;\hgam}= \ell_{k,p;\hgam[-p\eps^a/p^0]},
\label{Moninv}
\ee
where the parameter of the compensating spectral flow transformation is now $\hat\eps^a=p\eps^a/p^0$.
In principle, \eqref{Moninv} follows from the above results for the Heisenberg transformations and the commutation relations
\eqref{monS} and \eqref{commT}. One can prove it also by a direct computation in the way
which is completely analogous to the one for $\opT^{(1)}_{0,\eta^a}$,
except that one should use the reciprocity relation \eqref{reclaw} to express the Dedekind sum proving the cancelation of
the constant phase factor.

Note however that the transformation properties \eqref{Heisinv} and \eqref{Moninv} become quite different
once they are rewritten in terms of the transition functions \eqref{tranNS5all}.
Whereas these functions satisfy exactly the same constraint as \eqref{Heisinv}, the analogue of \eqref{Moninv} is a non-linear property
\be
\opT^{(1)}_{\eps^a,0}\cdot \Hij{\hgam[p\eps^a/p^0]}_{k,p}= \Hij{ \hgam}_{k,p}
+ \frac12\, \kappa_{abc} \epsilon^a \Tmn{b} \Tmn{c} + \frac12\, \kappa_{abc} \epsilon^a \epsilon^b \Tmn{c} \Tmn{0}
+ \frac16\,\kappa_{abc} \epsilon^a \epsilon^b \epsilon^c (\Tmn{0})^2,
\label{Moninv-tr}
\ee
where $\Tmn{\Lambda}$ was defined below \eqref{Sdualconstr}. This is indeed the right transformation law for the transition functions
under monodromies, which appears as a consequence of \eqref{monDc} and the gluing conditions \eqref{QKgluing}.
It illustrates once more the statement that isometries of a QK manifold are always realized in a linear way on
contact Hamiltonians, whereas their realization on transition functions may be highly non-linear.

\section{Derivatives of transition functions}
\label{ap-derfivebrane}

In this appendix we compute the combinations of the fivebrane transition functions appearing in the gluing conditions \eqref{QKgluing},
as these are the combinations entering the integrands of the integral equations determining the Darboux coordinates
on the twistor space and thereby the metric on the $\cM_H$.

Since we know the contact Hamiltonians generating fivebrane corrections,
the corresponding transition functions and their derivatives can be obtained by evaluating the action \eqref{newglue}.
To perform the calculations, we will need the following crucial observation:
for any homogeneous function $f_l(\xi)$ of degree $l$ one has
\be
\Xf{\hHij{\hgam}_{k,p}} \cdot \[ f_l(\xi+n) (\hkp)^m\]= -2\pi\I k (l+m)f_l(\xi+n)(\hkp)^{m+1}.
\label{homprop}
\ee
Due to this, one finds
\be
\begin{array}{rcll}
\Xf{\hHij{\hgam}_{k,p}}^m \cdot f_l(\xi+n) \hkp &=& 0 ,
\quad &{\rm if }\quad -(m+1)< l<0,
\\
\Xf{\hHij{\hgam}_{k,p}}^m \cdot f_l(\xi+n) \hkp &=&  (2\pi\I k)^m\frac{|l+1|!}{|l+m+1|!}\, f_l(\xi+n)(\hkp)^{m+1},
\quad &{\rm if }\quad l\le -(m+1),
\\
\Xf{\hHij{\hgam}_{k,p}}^m \cdot f_l(\xi+n) \hkp &=&  (-2\pi\I k)^m\frac{(l+m)!}{l!}\, f_l(\xi+n)(\hkp)^{m+1},
\quad & {\rm if }\quad l\ge0.
\end{array}
\ee
Another useful identity is
\be
\begin{split}
\Xf{\hHij{\hgam}_{k,p}}^m \,\cdot\, &\[ \p_{\xi^\Lambda}\tS_{k,p;\gamD{1}}\, \expe{\tS_{k,p;\gamD{1}}}\hkp \]
\\
&\, =
\(-\frac{4\pi^2 p^0 p^\Lambda\tilde q_\Lambda}{k(\xi^0+n^0)}\,\hkp\)^m
\( \p_{\xi^\Lambda}\tS_{k,p;\gamD{1}}-\frac{m\delta_{\Lambda}^0}{2\pi\I(\xi^0+n^0)}\)\expe{\tS_{k,p;\gamD{1}}}\hkp.
\end{split}
\ee
Using these properties, one computes\footnote{On the l.h.s. we use the same notations for patch indices
as in \eqref{Sdualconstr} and omit them on the r.h.s.}
\begin{subequations}
\bea
\Tkp{\Lambda}&\equiv&
\(\p_{\txii{\indrp}_\Lambda}- \xii{\indrp}^\Lambda \p_{\ai{\indrp}}\)\Hij{\hgam}_{k,p}
= -\( e^{\Xf{\hHij{\hgam}_{k,p}}}-1\)\cdot \xi^\Lambda
=
2\pi\I k \sum_{m=1}^\infty \frac{1}{m!}\,\Xf{\hHij{\hgam}_{k,p}}^{m-1} \cdot \(\xi^\Lambda+n^\Lambda\) \hHij{\hgam}_{k,p}
\nonumber\\
&=&
\frac{2\pi\I k\hHij{\hgam}_{k,p}\(\xi^\Lambda+n^\Lambda\)}{1+2\pi\I k\hHij{\hgam}_{k,p}},
\label{Tlambda}
\\
\tTkp{a}&\equiv&
\p_{\xii{\indlp}^a}\Hij{\hgam}_{k,p}= \( e^{\Xf{\hkp}}-1\)\cdot \txi_a
=2\pi\I  \sum_{m=1}^\infty \frac{1}{m!}\,\Xf{\hkp}^{m-1} \cdot
\(\p_{\xi^a}S_{k,p;\hgam}\)\hkp
\nonumber\\
&=&
 -\frac{2\pi\I  p^0 \hat q_a}{k(\xi^0 + n^0)}\, \hkp
-\frac{2\pi\I k \hkp}{1+2\pi\I k \hkp}\,\Fcl_a(\xi + n  )
\nonumber\\
&&
- \frac{(-1)^{\Nq_{k,p}(\hgam)}}{2\pi \I} \sum_{\gamD{1}\in\Gamma_{k,p;\hgam}}\bn_{\tilde q}\,
\tilde {q}_a \expe{\tS_{k,p;\gamD{1}}}
\[1 - e^{-4\pi^2\tilde{q}_\Lambda p^\Lambda \frac{p^0  \hkp}{k(\xi^0 + n^0)}} \],
\label{tTlambda-a}
\\
\tTkp{0}&\equiv&
\p_{\xii{\indlp}^0}\Hij{\hgam}_{k,p}= \( e^{\Xf{\hkp}}-1\)\cdot \txi_0
=  \sum_{m=1}^\infty \frac{1}{m!}\,\Xf{\hkp}^{m-1} \cdot
\((\xi^0+n^0)^{-1}+2\pi\I\p_{\xi^0}S_{k,p;\hgam}\)\hkp
\nonumber\\
&=&
 \frac{\hkp}{\xi^0+n^0}+\frac{2\pi\I p^0\(k\hat q_a(\xi^a+n^a)-p^0\hat q_0\)}{k^2(\xi^0+n^0)^2}\,\hkp
+\frac{2\pi^2 (p^0)^2\hat q_0}{k(\xi^0+n^0)^2}\, (\hkp)^2
\nonumber\\
&&
-\frac{2\pi\I k \hkp}{1+2\pi\I k \hkp}\,\Fcl_0(\xi + n  )
-\frac{(-1)^{\Nq_{k,p}(\hgam)}}{4\pi^2}  \sum_{\gamD{1}\in\Gamma_{k,p;\hgam}}\bn_{\tilde q}\,
\Biggl[\frac{k}{p^0}\, \cE\(-\frac{4\pi^2 p^0 p^\Lambda\tilde q_\Lambda}{k(\xi^0+n^0)}\,\hkp\)
\Biggr.
\nonumber\\
&&
\Biggl.\qquad\qquad
-2\pi\I\,\frac{\tilde q_0 p^0-k\tilde q_a\xi^a}{k(\xi^0+n^0)}
\(1 - e^{-4\pi^2\tilde{q}_\Lambda p^\Lambda \frac{p^0  \hkp}{k(\xi^0 + n^0)}} \) \Biggr]
\expe{\tS_{k,p;\gamD{1}}},
\label{tTlambda-0}
\eea
\bea
\tTkpa&\equiv &\(1-\xii{\indlp}^\Lambda\p_{\xii{\indlp}^\Lambda}\)\Hij{\hgam}_{k,p}= \( e^{\Xf{\hkp}}-1\)\cdot \alpha
\nonumber\\
&=& \sum_{m=1}^\infty \frac{1}{m!}\,\Xf{\hkp}^{m-1} \cdot \(\frac{n^0}{\xi^0+n^0}-2\pi\I \xi^\Lambda \p_{\xi^\Lambda}S_{k,p;\hgam}\)\hkp
\nonumber\\
&=& \frac{n^0\hkp}{\xi^0+n^0}
+2\pi\I\,\frac{(p^0)^2\hat q_0 \xi^0+p^0 k\hat q_a(\xi^a n^0-\xi^0 n^a)}{k^2(\xi^0+n^0)^2}\,\hkp
+\frac{2\pi^2(p^0)^2\hat q_0 n^0}{k(\xi^0+n^0)^2}\,\hkp
\nonumber\\
&&+\(1-\(1+2\pi\I k \hkp\)^{-2}\)F(\xi+n)-\frac{2\pi\I k \hkp}{1+2\pi\I k \hkp}\, n^\Lambda F_\Lambda(\xi+n)
\nonumber\\
&&-\frac{(-1)^{\Nq_{k,p}(\hgam)}}{4\pi^2}  \sum_{\gamD{1}\in\Gamma_{k,p;\hgam}}\bn_{\tilde q} \,
\Biggl[\frac{p}{p^0}\, \cE\(-\frac{4\pi^2 p^0 p^\Lambda\tilde q_\Lambda}{k(\xi^0+n^0)}\,\hkp\)
\Biggr.
\nonumber\\
&&
\Biggr.\qquad\qquad
+2\pi\I\,\frac{ p^0\tilde q_0\xi^0  +p\tilde q_a\xi^a}{k(\xi^0+n^0)}
\(1 - e^{-4\pi^2\tilde{q}_\Lambda p^\Lambda \frac{p^0  \hkp}{k(\xi^0 + n^0)}} \) \Biggl]
\expe{\tS_{k,p;\gamD{1}}}.
\label{Talpha}
\eea
\label{evalHk}
\end{subequations}
It is straightforward to check that these results lead to the expression for the transition functions $\Hij{\hgam}_{k,p}$
given in \eqref{tranNS5all}.

\providecommand{\href}[2]{#2}\begingroup\raggedright\endgroup


\end{document}